# A Technique for Multi-User MIMO using Spatial Channel Model for out-door environments

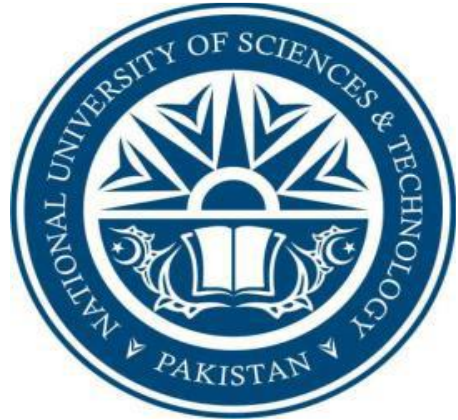

Author

SYED WAQAS HAIDER SHAH

NUST201464406MCEME35014F

Supervisor

DR. SHAHZAD AMIN SHEIKH

DEPARTMENT OF ELECTRICAL ENGINEERING

COLLEGE OF ELECTRICAL & MECHANICAL ENGINEERING

NATIONAL UNIVERSITY OF SCIENCES AND TECHNOLOGY

ISLAMABAD

September, 2016

# A Technique for Multi-User MIMO using Spatial Channel Model for out-door environments

Author

SYED WAQAS HAIDER SHAH

NUST201464406MCEME35014F

A thesis submitted in partial fulfillment of the requirements for the degree of

MS Electrical Engineering

Thesis Supervisor:

DR. SHAHZAD AMIN SHEIKH

Thesis Supervisor's Signature: \_\_\_\_\_\_\_\_\_\_\_\_\_\_\_\_\_\_\_\_\_\_\_\_\_\_\_\_\_\_\_\_\_\_\_\_\_\_\_\_

DEPARTMENT OF ELECTRICAL ENGINEERING

COLLEGE OF ELECTRICAL & MECHANICAL ENGINEERING

NATIONAL UNIVERSITY OF SCIENCES AND TECHNOLOGY, ISLAMABAD

September, 2016

# Declaration

I certify that this research work titled "*A Technique for Multi-User MIMO using Spatial Channel Model for out-door environments"* is my own work. This work has never been presented anywhere else before and the data taken from different sources is properly referred.

<div style="text-align:right">

Signature of Student

SYED WAQAS HAIDER SHAH

NUST201464406MCEME35014F

</div>

# Language Correctness Certificate

This thesis report has been read by an expert and is free from spelling, syntax, and grammatical errors. The thesis is written according to the format provided by the university.

<div style="text-align: right;">
Signature of Student
SYED WAQAS HAIDER SHAH
NUST201464406MCEME35014F

Signature of Supervisor
</div>

# Copyright Statement

- Copyright in content of this proposition rests with the understudy creator. Duplicates (by any procedure) either in full, or of concentrates, might be made just as per guidelines given by the creator and stopped in the Library of NUST College of E&ME. Subtle elements might be gotten by the Librarian. This page must frame part of any such duplicates made. Further duplicates (by any procedure) may not be made without the consent (in composing) of the creator.

- The responsibility for protected innovation rights which might be depicted in this proposition is vested in NUST College of E&ME, subject to any earlier consent in actuality, and may not be made accessible for use by outsiders without the composed authorization of the College of E&ME, which will recommend the terms and states of any such understanding.

- Further data on the conditions under which revelations and misuse may occur is accessible from the Library of NUST College of E&ME, Rawalpindi.

# Acknowledgements

I would like to thank Allah (SWT) who has always showered his countless blessings on me.

A special thanks to my supervisor Dr. Shahzad Amin Sheikh and Brig. Dr. Khalid Iqbal for his stupendous supervision, and a very consistent encouragement throughout the course of this thesis. I wouldn't have been here today without his encouragement.

I would also like to thank Dr. Fahad Mumtaz Malik and Dr. Usman Ali for being on my thesis Guidance and Examination Committee and for guidance. A very special thanks from the bottom of my heart to my favorite teacher Dr. Shahzad Amin Sheikh who has always been with me through thick and thin throughout my entire journey at EME.

I am also thankful to Maj. Muhammad Latif, Sajid Bshir, Mudassar Hussain, Talal Riaz, Jawad Javaid, Muhammad Haras and Muhammad Imran for their support and encouragement from the very beginning of my master's program.

Finally, thanks to my parents who always encouraged me whenever I was stuck or confused with the work.

*To Mom and Dad,*

*Who always picked me up on time*

*And encouraged me to go on every adventure*

*Especially this one*

# ABSTRACT

Any wireless communication system needs to specify a propagation channel model which acts as basis for performance evaluation and comparison. Spatial channel models can be divided into deterministic i.e ray tracing, measurement based which is based on channel information and geometry based stochastic channel models which are based on assumption that directional structure of channel can be modelled by last interaction between physical objects and electromagnetic waves, before waves reach the base or mobile station. Multi user double directional channel model (MDDCM) is a geometry based channel model which is used to calculate the double directional channel information in cellular system with MIMO mobile station and MIMO base station.

In this research phase firstly, Multi User Multiple Input Multiple Output (MU-MIMO) spatial channel model has been implemented for different outdoor environments Urban Micro and Urban Macro using MATLAB for finding various parameters like angle of arrival of the user, user direction and the distance between user and access point (AP). Secondly coded (Multiple Input Multiple Output-Orthogonal Frequency Division Multiplexing) MIMO-OFDM system has been implemented using multipath Rayleigh faded channel and realistic Spatial Channel Model. Different BER improvement techniques are used such as, Viterbi-decoder, Time and Frequency inter-leaving. Multi-channel diversity is also observed by using multiple antennas at transmitting and receiving end [(2x2) and (2x4)] on both the channels. Effect of different modulation techniques on BER performance is also observed.

# Table of Contents

# List of Figures

# List of Tables

# Acronyms

| | |
|---|---|
| SCM | Spatial Channel Model |
| MIMO | Multiple Input Multiple Output |
| MU-MIIMO | Multi User Multiple Input Multiple Output |
| OFDM | Orthogonal Frequency Division Multiplexing |
| AP | Access Point |
| STBC | Space Time Coding |
| WINNER | the Wireless World Initiative New Radio |
| BER | Bit Error Rate |
| SNR | Signal to Noise Ratio |
| AoA | Angle of Arrival |
| AoD | Angle of Departure |
| RFC | Rayleigh Fading Channel |
| BS | Base Station |
| MS | Mobile Station |

# CHAPTER 1

# INTRODUCTION

## 1.1 Introduction

There are different classification of models such as physical or mathematical. Mathematical equations and symbolic notations are used to represent a system in a mathematical model. A simulation models are specific type of mathematical models of a system which can be classified further as being discrete or continuous, deterministic or stochastic and static or dynamic. The static simulation models from time to time called as Monte-carlo simulation represent a system at particular point in time, whereas a dynamic simulation models represent a system as there transformation according to time. Simulations that contain no random variables are called ''Deterministic'', whereas a stochastic simulation has one or more random variables as inputs.

Any wireless communication system needs to specify a propagation channel model which acts as basis for performance evaluation and comparison. Spatial channel models can be distributed into deterministic i.e ray tracing, measurement based which is based on geometry built and channel information defined stochastic channel models which are created on hypothesis that directional structure of channel can be modelled by last communication between electromagnetic waves and physical objects, before waves arrives at mobile station or base. Multi user double directional channel model (MDDCM) is a geometry based channel model which is used to calculate the double directional channel information in cellular system with MIMO mobile station and MIMO base station.

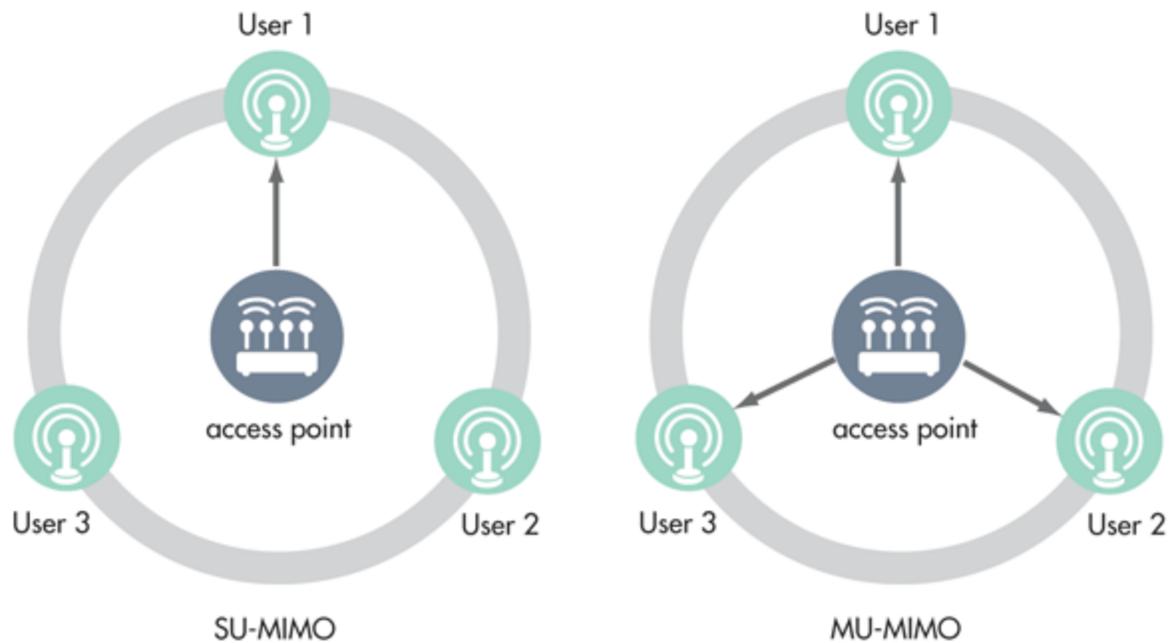

Figure1.1: SU-MIMO vs MU-MIMO

When electromagnetic waves propagate through a channel, they experience delays and losses due to the channel. If a line of sight (LOS) path exists between transmitter and receiver, the LOS multipath components will arrive first and additional physical paths throughout the environment cause remainder of the multi path components to arrive later in time. A signal propagation through any wireless channel can be affected by multiple phenomena such as diffraction, refraction, reflection and scattering of the radiated energy. These phenomenon can cause path-loss and fading in propagation. The fading is further divided into fast fading also known as multipath fading or Rayleigh fading and slow fading. Fast fading occurs due to multipath, where multiple reflected signals are summed up to a received signal, arriving with different phases and amplitudes at slightly different time. While on the other hand the shadowing effects of terrain, trees and buildings causes the slow fading. In case of NLOS path the envelope of received signal is Rayleigh distributed and is called Rayleigh fading. Whereas in case of dominant LOS path, the envelope will have a Rician distribution with less severe fading known as Rician fading.

## 1.2 Literature Review

Multi Input Multi Output (MIMO) technology has been standardized for wireless LAN, 3G mobile phone networks and 4G mobile phone networks to achieve higher data rates. The MIMO technique is the use of multiple antennas at both transmitter and receiver to improve data rate performance while beam-forming increases the signal to noise ratio (SNR) resulting in large coverage area. [1] Later, in order to increase the data rate even further, George Jongern proposed a technique for SU-MIMO known as Mode switching between SU-MIMO and MU-MIMO for LTE networks. [2-3]

The more advanced Multi User Multiple Input Multiple Output (MU-MIMO) technique is used to improve mobile broadband services and to support wider transmission bandwidths, as multiple users can transmit data simultaneously using multiple antennas at both receiving and transmitting end. [4] In theory MU-MIMO can also give throughput gains that have the ability to scale linearly with the number of antennas. [5] MU-MIMO is already supported in LTE release 8 via transmission mode 5 [TM5]. In LTE technology, the specifications for downlink and uplink rates are 300Mbit/s and 75Mbit/s respectively. [6-7] Experimental evaluation has already been done, and verified that antenna inter-element spacing for MU-MIMO is beneficial for outdoor environments. [8-9]

Recently Orthogonal Frequency Division Multiplexing MIMO known as MIMO-OFM for WIMAX has been widely accepted as the alternative to cellular standards. [10] WIMAX is based on the 802.1be standard and it uses MIMO-OFDM to further enhance the delivery speed up to 138Mbit/s. The more advanced version of it, 802.16m standard enables download speed up to 1Gbit/s. [11]

While on the other hand frequency and time inter-leaving is a technique used for making forward error correction more robust with respect to burst errors. [12]Therefore, MIMO processing techniques, such as space time coding, spatial multiplexing and diversity schemes have gained much attention.[13] MIMO diversity techniques and space time coding can be used for improvement of signal to noise ratio (SNR), whereas spatial multiplexing can be used for the improvement of data rates. [14-16] Overall MIMO processing techniques appears very promising for future wireless systems.

## 1.3   Contributions

It is obvious from the literature review that valuable research work has been submitted on MIMO-OFDM but we have implemented a technique for MU-MIMO using Spatial Channel Model on various outdoor environments. MU-MIMO Spatial channel model has been used to indicate the angle of different users operating from different locations in addition to following:

- Angle of Arrival of signal
- Direction of movement of User (MS).
- Distance between User (MS) and AP (Access Point (BS)).

After simulation of MU-MIMO on SCM, we have implemented coded OFDM-MIMO technique on Rayleigh faded channel and realistic Spatial channel model (SCM). Different BER improvement techniques are used such as, Viterbi-decoder, Time and Frequency inter-leaving. Multichannel diversity is also observed by using multiple antennas at transmitting and receiving end [(2x2) and (2x4)] on both the channels. Effect of different modulation techniques on BER performance is also observed.

# CHAPTER 2

# SPATIAL CHANNEL MODEL

## 2.1 Development of Spatial Channel Model

For simulation and design of smart antenna systems, spatial channel model is needed that reflects the measured characteristics of a mobile radio channel. There should be a specific propagation channel model which plays a role as a performance evaluator and comparator. Spatial channel model (SCM) is called geometric or ray based model which is based on stochastic modelling of scatterers. In Spatial Channel Model these environments such as, urban macro and urban micro are considered. Urban micro is also defined as NLOS and LOS propagation. Every scenario is being given fixed number of paths which can be modified in channel parameter configuration function and every path has further separated with twenty (20) spatially sub paths. This channel model is used to generate the matrices for desired number of links by using different parameters in the input structures, Such as channel configuration parameters, antenna-parameters and link-parameters. This SCM channel model gives output the MU-MIMO channel matrices while having the input of link-parameter, antenna-parameter and channel configuration-parameter. Channel impulse response for pre-defined number of links is given by a multi-dimensional array output.

## 2.2 Environments Considered for MU-MIMO SCM

Four environments considered in MU-MIMO spatial channel model are as under.

### 2.2.1 Urban Environments

An urban area is described as heavily built up area within a city. Tall buildings along streets act as reflectors of radio waves and LOS path normally does not exist because of shadowing of nearby buildings. Both the base station and mobile antennas presumably use an Omni-directional antenna.

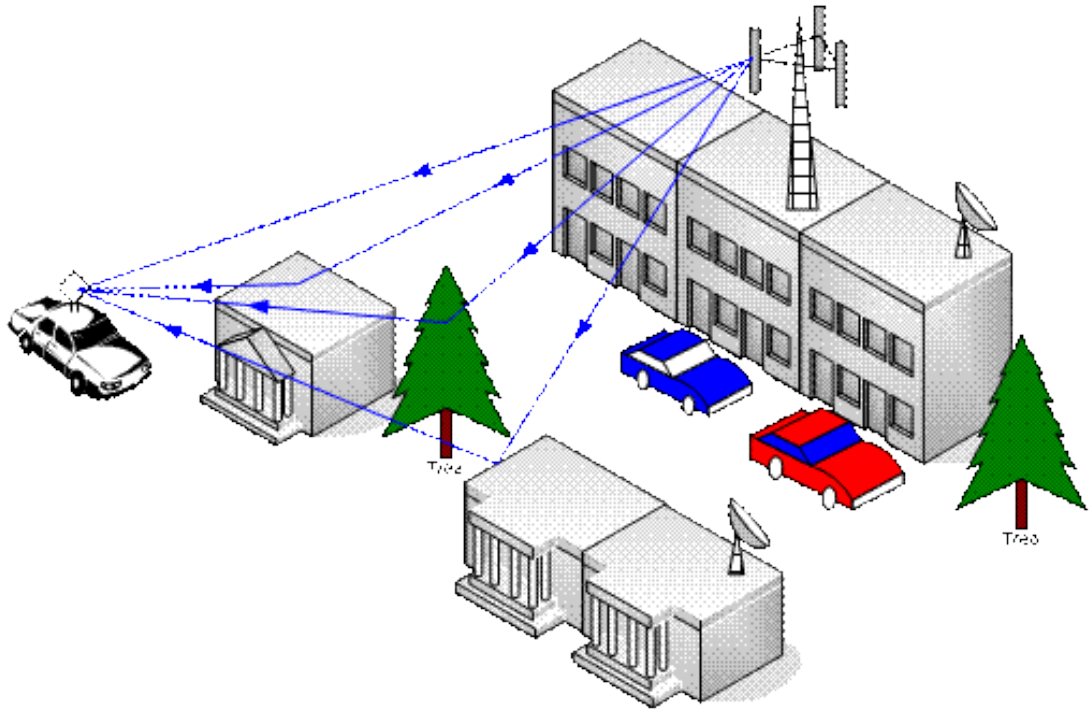

Figure2.1: Urban Environment

## 2.2.2 Sub-Urban Environments

A sub-urban area is described as a less built up outskirts of a city. These areas may be open farmlands and there may also be some visible mountains off in the distance. In sub-urban areas nearby buildings cause most of the multi-path with small time delays, but the large scatterers such as large buildings and mountains, generate significant multi-path components with large time delays.

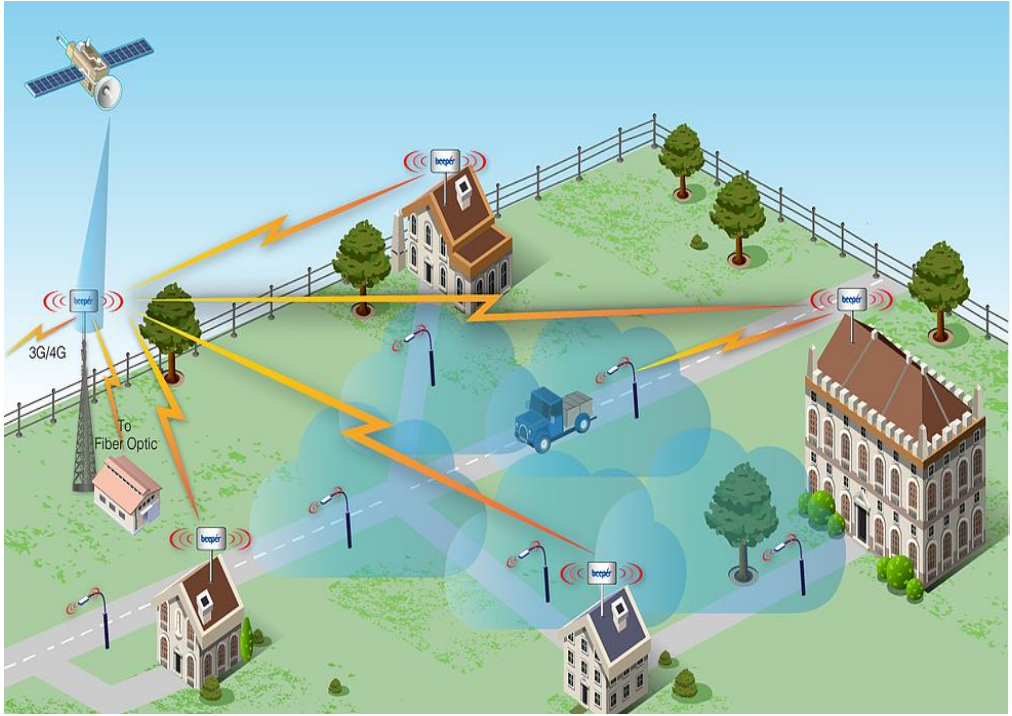

Figure2.2: Sub-Urban Environment

### 2.2.3 Macro cell Environments

In case of macro cell environment, scatterers surrounding MS are at same height or higher than MS, hence BS antenna is placed above scatterers.

### 2.2.4 Micro cell Environment

In micro cell environment, BS antenna is placed at almost same height as the objects around it. In micro cell environment the scattering spread of received signal at BS is greater than that of macro cell environment and delay spread is less due to smaller coverage area.

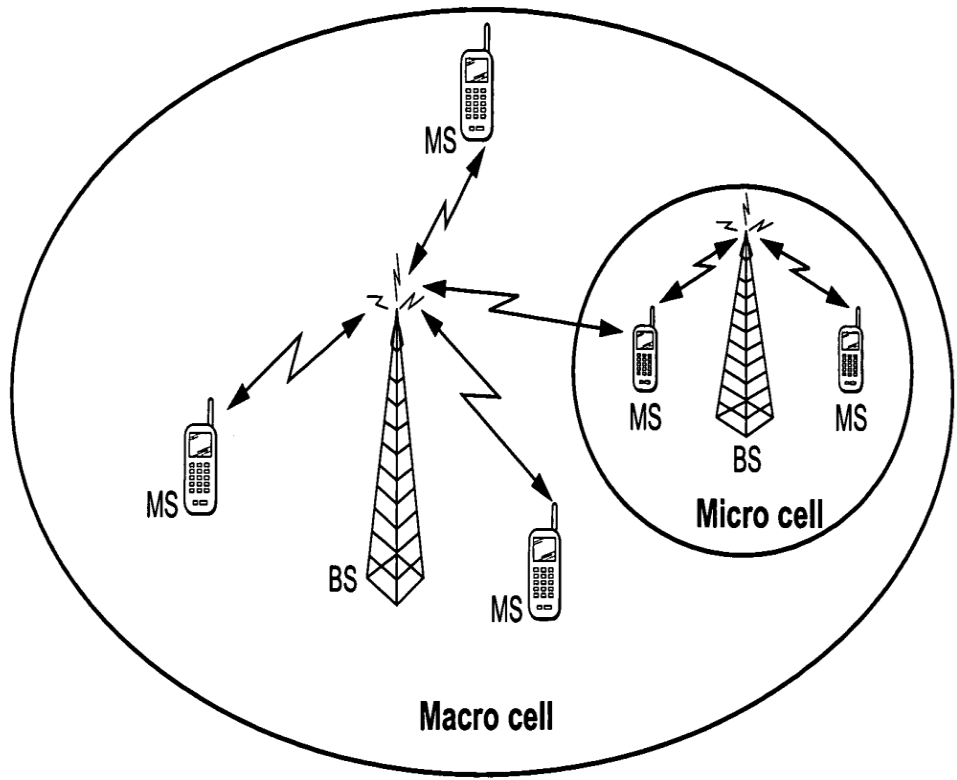

Figure2.3: Macro and Micro Cell Environment

## 2.3 Syntax for the use of SCM

The full syntax for SCM is given as [CHAN, OUTCOME, DELAY CALCULATION] = Spatial_channel (CHANPARSET, ANTPARSET, LINKPARSET) whereas,

- CHANPARSET, ANTPARSET, LINKPARSET are generated as MATLAB structures.
- The first output CHANPARSET is a FIVE Dimensional (5D) array containing the Multi Input Multi Output Spatial Channel matrices for all links over a specified number of time samples.
- The second output argument is a MATLAB structure and elements of this structure [OUTCOME] contains the information of delays, power of each path, angle of departure, angle of arrival of all twenty (20) spatially separated sub paths and its phases, path losses, shadow fading and time difference (delta-t), a vector which defines time sampling interval for all links.

- The third output DELAY CALCULATION defines delays of multipath for every link. These delays are given in seconds.

## 2.4 SIMULATION OF MU-MIMO SCM

An Uplink case is being simulated using Multi-User Multi Input Multi Output (MU-MIMO) spatial channel model (SCM) for calculating angle of arrival (AoA) at AP from user which is indicated by a broad beam in the direction of user. The channel experiences fast fading in each simulation run according to the motion of user. To determine the user's direction where to transmit the AP uses the schedulers which are being generated from channel state information (CSI) which is fed back from user to AP. The channel matrix co-efficient are being generated by using Spatial_channel.m which are then gives us the information of angle of arrival (AoA) of the user. To set the parameters for input structures like links, antenna and SCM model we use LINKPARSET, ANTPARSET and CHANPARSET respectively.

### 2.4.1 SCM Parameter Set

Various parameters of this input structure are defined in Table 2.1 below.

| NumAPElements | Number of antenna array elements used in access point (AP) |
|---|---|
| NumUserElements | Number of antenna array elements used in user station. |
| Environment | Scenarios which could be, urban micro or urban macro. |
| Sample density Value | It states the number of samples per half wavelength. Also defined as channel's sampling interval. As the Doppler analysis is required so a value greater than one '1' i.e 3 in this case is selected. |
| APUrban-MacroAS | Average Angle Spread (Mean) of User: 80º and 150º are selected which are only possible values for Urban-macro environment. |
| No_Paths | Total number of paths available which can be changeable according to scenario. |
| S_paths_per_path | Total no. of sub-paths available in each path. 20 sub-paths are selected for SCM. Table 2.2 is given for the offset AoD/AoA for every sub path. |
| CF | Central frequency (2.0 GHz) which can affects the time sampling interval and path loss. |
| Chan-Options | SCM channel Options which can be urban canyon, polarized, LOS or none. All of these are mutually exclusive options. |

TABLE 2.1. SCM Parameter Set

| S-path No. (*n*) | 2º Angle Spread at AP (Urban-Macro cell) $\Delta_{t,n,AoD}$ (degrees) | 5º Angle Spread at AP (Urban-Microcell) $\Delta_{t,n,AoD}$ (degrees) | 35º Angle Spread at user station $\Delta_{t,n,AoA}$ (degrees) |
|---|---|---|---|
| 1&2 | ± 0.0784 | ± 0.3012 | ± 1.4985 |
| 3&4 | ± 0.3197 | ± 0.7573 | ± 5.1425 |
| 5&6 | ± 0.5013 | ± 1.1989 | ± 9.0190 |
| 7&8 | ± 0.8014 | ± 1.9147 | ± 12.8045 |
| 9&10 | ± 1.1348 | ± 2.6524 | ± 15.8562 |
| 11&12 | ± 1.2945 | ± 3.4572 | ± 22.8766 |
| 13&14 | ± 1.8541 | ± 4.5142 | ± 31.0487 |
| 15&16 | ± 2.3416 | ± 5.6942 | ± 39.5124 |
| 17&18 | ± 2.9984 | ± 7.4265 | ± 51.2375 |
| 19&20 | ± 4.2132 | ± 10.8754 | ± 74.5423 |

TABLE 2.2. S-path offsets of AoA and AoD

### 2.4.2 Antenna Parameter Set (ANTPARSET)

This is also being used for defining the input antenna parameter configuration for MU-MIMO SCM. The identical behavior of antenna pattern is not necessary; it only supports the linear arrays in this case. The main fields of the antenna parameter set (ANTPARSET) are given in Table 2.3.

| | |
|---|---|
| AP-G-Pattern | This is an argument which defines Access Point gain pattern. All the elements have uniform and identical gain so the value is set to '1'. |

| AP-Azimuth-angles | This input argument is a vector which contains the information of Azimuth angles for the field pattern values of Access Point (AP). Its value is set in the range of –π (-180) to +π (+180). |
|---|---|
| AP-Elem-Pos | It defines the Access Point's position of linear antenna array in wavelength, 0.5 is selected as a uniform spacing between the elements. |
| User-G-Pattern | This is an argument which defines User (Mobile station) gain pattern. All the elements have uniform and identical gain so the value is set to '1'. |
| User-Elem-Pos | It defines the User's position of linear antenna array in wavelength, 0.5 is selected as a uniform spacing between the elements. |
| User-Azimuth-Angles | This input argument is a vector which contains the information of Azimuth angles for the field pattern values of User. Its value is set in the range of –π (-180) to +π (+180). |

TABLE2.3. Antenna Parameter Set

### 2.4.3 Link Parameter Set (LINKPARSET)

This is also being used for defining the input Link parameter configuration for MU-MIMO SCM. Every parameter is a vector of length 'N', where N is the no. of links. The main fields of the antenna parameter set (LINKPARSET) are given in Table 2.4.

| AP-USER-Distance | This input argument is a vector which contains the information of the distance between User and AP, every user is set to be 35m to 500m away from the AP in this case. |
|---|---|
| $\Theta_{AP}$ | It contains the angle of arrival of the signals for AP in degree. |
| $\Theta_{User}$ | It contains the angles of User in degree. |
| $V_{User}$ | Velocity of the user in meter/sec. (m/s) |
| User-Direct | It contains the information of direction of the User with respect to Broadside of User antenna array. |
| User-Height | Height of the user from the ground surface, it is set to 1.5m. |
| AP-Height | Height of AP from the ground surface, it is set to 32m. |

| User-No | It is a vector of 1….N, N is the number of links available. It defines the number of users available in each simulation run. |

TABLE 2.4. Link Parameter Set

### 2.4.4 Output Argument

The output argument ''W'' is a FIVE DIMENSIONAL (5D) array and is defined as under.

Size (W) = [L M N K S]

Whereas,

L = Number of antenna elements available for Access Point (AP).

M = Number of antenna elements available for User.

N = Number of links

K = Total number of paths available for transmission.

S = Total number of time samples are generated per path.

These parameters are used for the generation of the channel co-efficient. For an ''L'' elements linear AP antenna array and ''M'' elements linear User antenna array, LxM matrix of complex amplitudes will give the information of channel co-efficient for 'K' no. of paths. The channel matrix for kth path (n = 1………k) is denoted as Wk(t). Movement of User can cause fast fading in complex amplitudes so it becomes the function of time't'.

## 2.5 Generation of Channel Matrix

It takes three simple steps for the generation of Channel Matrix.

i. In first step it is required to define the environment as described above.

ii. In second step, need to acquire the parameters for particular environment

iii. In third step, Generation of the channel co-efficient based on the parameters calculated in second step.

The (l,m)th component (l = 1………L; m =1……..M) of Wk(t) is given by:

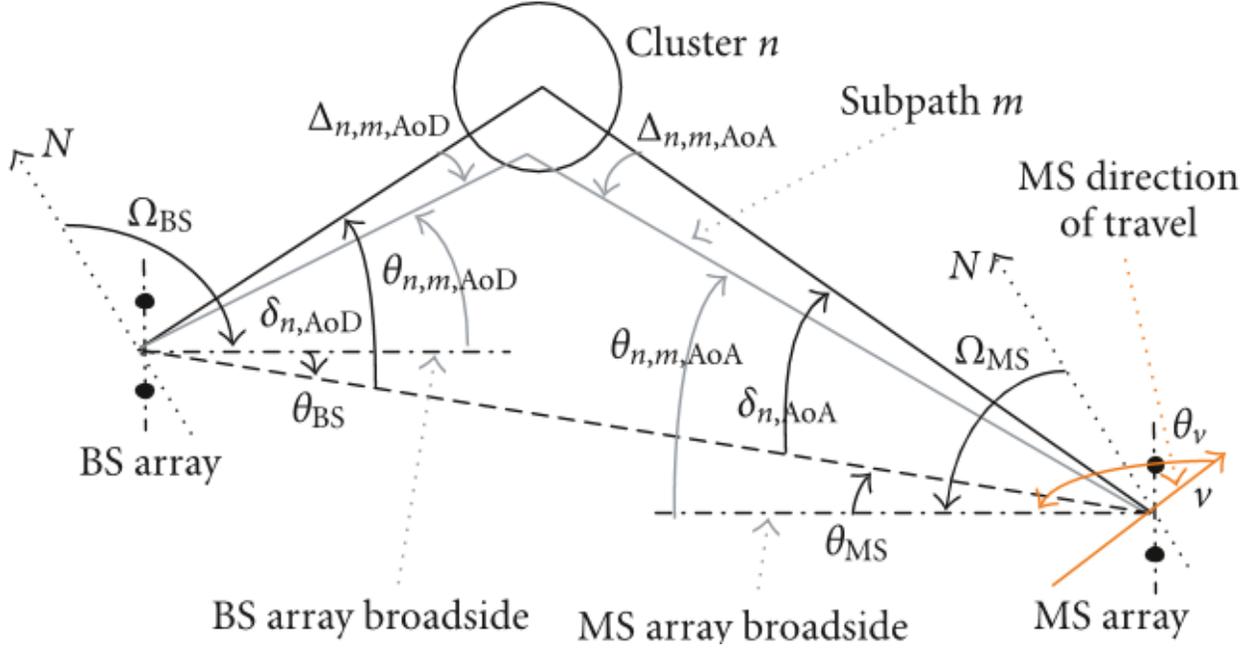

Figure2.4: SCM Channel Matrix

$$h_{u,s,n}(t) = \sqrt{\frac{P_n \sigma_{SF}}{S}} \sum_{m=1}^{M} \begin{bmatrix} \sqrt{G_{BS}(\theta_{t,n,AoD})} \exp(j[kd_s \sin(\theta_{t,n,AoD}) + \phi_{n,m}]) \times \\ \sqrt{G_{MS}(\theta_{t,n,AoA})} \exp(jkd_u \sin(\theta_{t,n,AoA})) \times \\ \exp(jk\|v\|\cos(\theta_{t,n,AoA} - \theta_v)t) \end{bmatrix} \quad (1)$$

Whereas,

$P_n$ = Power of $n^{th}$ path.

S = Total no. of sub-paths available in every path

$\sigma_{SF}$ = lognormal shadow fading.

$\theta_{t,n,AoD}$ = Departure Angle for $m^{th}$ sub_path of $n^{th}$ path.

$\theta_{t,n,AoA}$ = Angle of arrival for $m^{th}$ sub_path of $n^{th}$ path.

$G_{BS}(\theta_{t,n,AoD})$ = is the BS antenna gain of each array element

$G_{MS}(\theta_{t,n,AoA})$ = is the MS antenna gain of each array element

j = it is the square root of –1

k = 2Π/λ, where λ is the wavelength in meters

$d_l$ = distance between reference element and AP antenna element. Distance is in meters.

$d_m$ = distance between reference element and User antenna element. Distance is in meters.

$\phi_{n,m}$ = phase angle of $m^{th}$ sub_path of the $n^{th}$ path.

||v|| = Magnitude of User velocity vector

$\theta_v$ = Angle of User velocity vector

## 2.6 Simulation Results

MU-MIMO Spatial channel model has been used to indicate the angle of different users operating from different locations in addition to following:

a. Angle of Arrival of signal

b. Direction of movement of User.

C. Distance between User and AP.

### 2.6.1 Case 1

A model of a simple 1xtransmitter and 1xreceiver in urban-microenvironment using the SCM model considering design parameters defined in Table 5 is implemented in MATLAB. The resultant value of angle of arrival in Linear, Polar and MUSIC plots are shown in figure 2.5.

| Links (number of links/users) | 1 |
|---|---|
| Paths (number of Paths) | 1 |
| n_max (number of channel samples generated per Link) (impulse response matrices) | 10 |
| scmpar.NumAPElements (antenna elements in the AP antenna array) | 8 |
| scmpar.NumUserElements (antenna elements in the User antenna array) | 2 |
| scmpar.ScmOptions (Switches on the line of sight option) | "LOS" |
| User Velocity (Velocity of mobile) | 5m/sec |
| User Height (Height of user antenna elements) | 1.5m |
| AP Height (Height of AP antenna elements) | 32m |
| UserNumber (Number of Mobile Users) | 1 |
| D (inter-element spacing) | 0.5m |
| λ (wave length) | d/2 |

TABLE 2.5. Case-01 Design Parameters

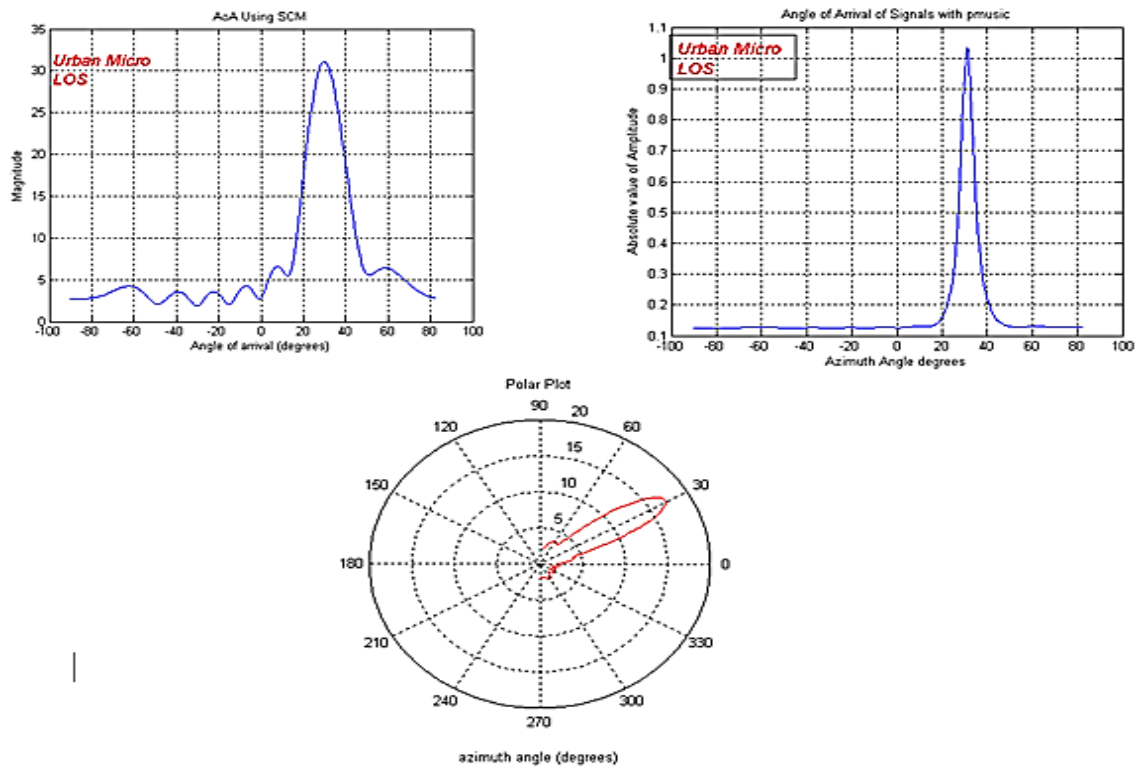

FIGURE 2.5. AOA of 1x Mobile User

### 2.6.2 Case 2

A model of a simple 1xtransmitter and 3x receivers in 'urban-micro' environment using the SCM model considering design parameters define in Table 6 is implemented in MATLAB. The resultant values of AoA using both Linear and Polar plot is shown in Figure2.6.

| | |
|---|---|
| Links (number of links/users) | 3 |
| Paths (number of Paths) | 3 |
| n_max<br>(number of channel samples generated per Link)<br>(impulse response matrices) | 10 |
| scmpar.NumAPElements (total antenna elements in the AP antenna array) | 8 |
| scmpar.NumUserElements (total antenna elements in the User antenna array) | 2 |
| scmpar.ScmOptions (Switches on the line of sight option) | "LOS" |
| UserVelocity (Velocity of mobile) | 5m/sec |
| UserHeight (Height of User antenna elements) | 1.5m |
| AP-Height (Height of AP antenna elements) | 32m |
| User-No (Number of Mobile Users) | [1 2 3] |
| d (inter-element spacing) | 0.5m |
| $\lambda$ (wave length) | d/2 |

TABLE 2.6. Case 2 Design parameters

The resultant output values calculated in the link-parameter structure are as under:

User-AP Distance = 349.2884m, 422.5138m and 422.8360m respectively for all users.

Angles of Arrival = -40o, 20o and 40o

User-Direction = 109.7538o, 147.0231o and -96.5180o all users

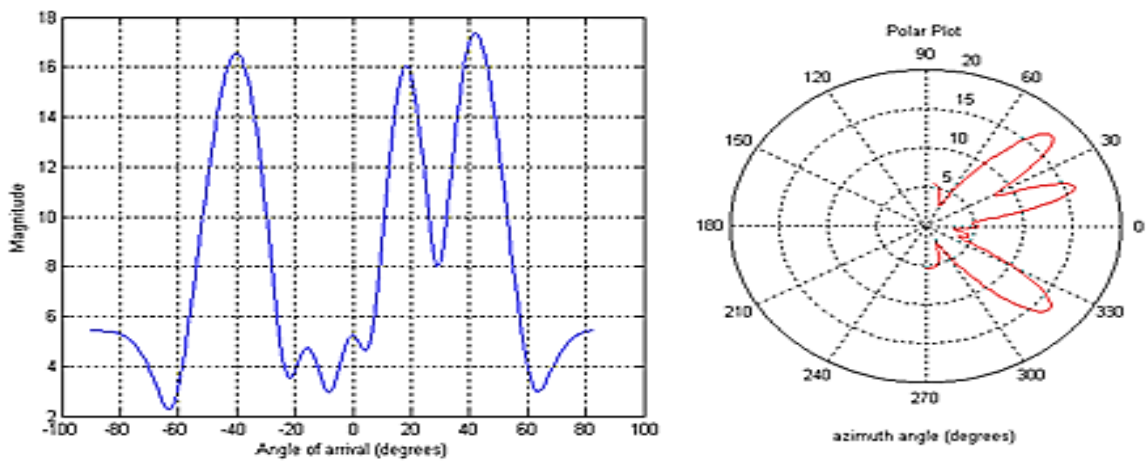

FIGURE 2.6. AoA of 3x Mobile Users

### 2.6.3 Case 3

A model of a simple 1x transmitter and 2x receivers in urban-macro environment using the SCM model considering design parameters defined in Tabl3 7 is implemented in MATLAB. The resultant values of Angles of Arrival using both Linear and Polar plot is shown in Figure 2.7.

| | |
|---|---|
| Links   (number of links/users) | 2 |
| Paths  (number of Paths) | 2 |
| n_max (number of channel samples generated per Link) (impulse response matrices) | 10 |
| scmpar.NumAPElements   (total antenna elements in the AP antenna array) | 8 |
| scmpar.NumUserElements  (total antenna elements in the User antenna array) | 2 |
| scmpar.ScmOptions  (Switches on the line of sight option) | "LOS" |
| User-Velocity  (Velocity of mobile) | 5m/sec |
| User-Height  (Height of User antenna elements) | 1.5m |
| AP-Height  (Height of AP antenna elements) | 32m |
| User-No  (Number of Mobile Users) | [1 2] |
| D  (inter-element spacing) | 0.5m |
| $\lambda$  (wave length) | d/2 |

TABLE 2.7.Case-03 Design Parameters

The resultant output values calculated in the Link-parameter structure are as under:

User-AP Distance=498.5580m and 374.4319m all users.

Angles of Arrival =  -40o and 40o

User-Direction = 104.8044o and 113.3827o  all users

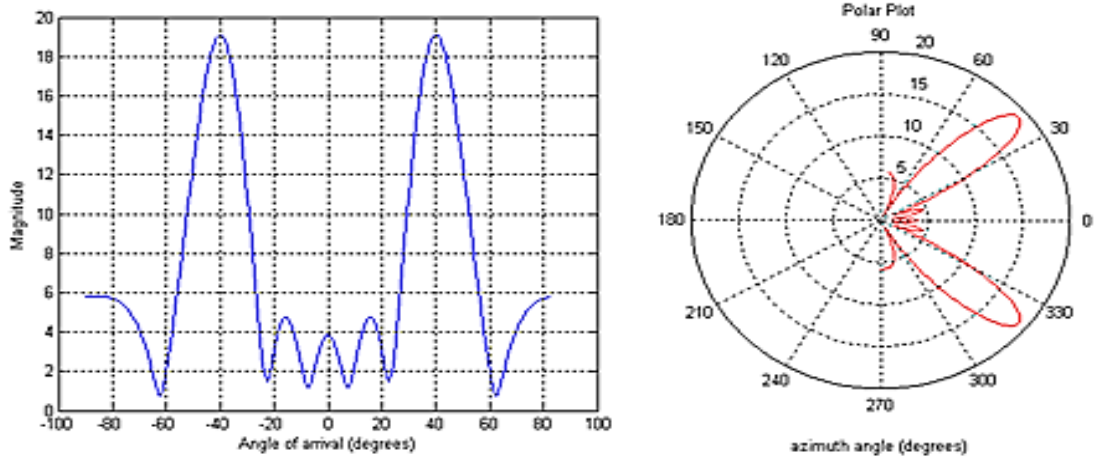

FIGURE 2.7: AoA's of 2x mobile users

Complete simulation results of simulation are given in Table 2.8. In this table a complete comparison of Angle of Arrival of the user, distance between User and Access Point (AP) and User direction of the movements are given for every user.

| No of Users | Input Parameters | | | | | | | | | Output Results | | | | | | | | |
|---|---|---|---|---|---|---|---|---|---|---|---|---|---|---|---|---|---|---|
| | antenna elements | | Scenarios | | | Option | | | Urban Scenario | MS velocity | Channel samples/link | AoAs (degrees) | | | MS-BS Distance (meters) | | | MS Direction (degrees) | | |
| | BS | MS | S-nr. benchmark scenario | Urban benchmark | Urban n-micro | LOS | None | | | | | 1 | 2 | 3 | 1 | 2 | 3 | 1 | 2 | 3 |
| 1 | 8 | 2 | - | - | Y | Y | - | - | 5 | 10 | 30 | - | - | 174.33 | - | - | -157.3 | - | - |
| 2 | 8 | 2 | - | Y | - | Y | - | - | 5 | 10 | -40 | 40 | - | 498 | 375 | - | 105.00 | 113.0 | |
| 3 | 8 | 2 | - | - | Y | Y | - | - | 5 | 10 | -40 | 20 | 40 | 349 | 422 | 423 | 109.0 | 147.0 | -96.0 |

TABLE 2.8. Simulation Results

# CHAPTER 3

# Multi Input Multi Output Orthogonal Frequency Division Multiplexing (MIMO-OFDM)

## 3.1 Introduction

In a real-time transmission scenario, there is mostly no line-of-sight transmission between the transmitting and receiving terminals and the channel conditions are time-varying which result in different phenomenon being applied to the signal such as reflection, refraction, diffraction and scattering along with distortion in amplitude and phase. Thus, the transmitted signal undergoes multipath propagation and multiple copies of signal are received at the receiver each with different time-delays and undergoing different magnitudes of distortions. These distortions are time-varying and thus the received signal levels are also time-varying. In addition to static hurdles, there are moving hurdles along with the possibility that transmitter or/and receiver are in motion. This relative motion between the two ends gives rise to Doppler's shift to the received signal and subsequently Inter-Carrier Interference (ICI) occurs, owing to the shift in the frequency band of the signal. To overcome these problems associated with the distortion of signal, a special technique called Orthogonal Frequency Division Multiplexing (OFDM) is adopted. OFDM scheme breaks down the spectrum into small sub-channels called subcarriers. The data symbols are then transmitted over these subcarriers by modulating over these subcarriers. There are several sub-carriers, depending upon the total available bandwidth, each with a carrier spacing selected meticulously such that it is orthogonal to all other subcarriers. The most obvious advantage of this division into sub-channels is that a highly frequency selective channel, in effect, behaves uniformly over the individual subcarrier, thus many flat-fading channels are achieved. Moreover, since the subcarriers are orthogonal to each other, there is lesser probability of Inter-Symbol Interference with a relatively higher bandwidth efficiency. At the receiver, the data symbols can be stripped off the orthogonal subcarriers using similar correlation techniques.

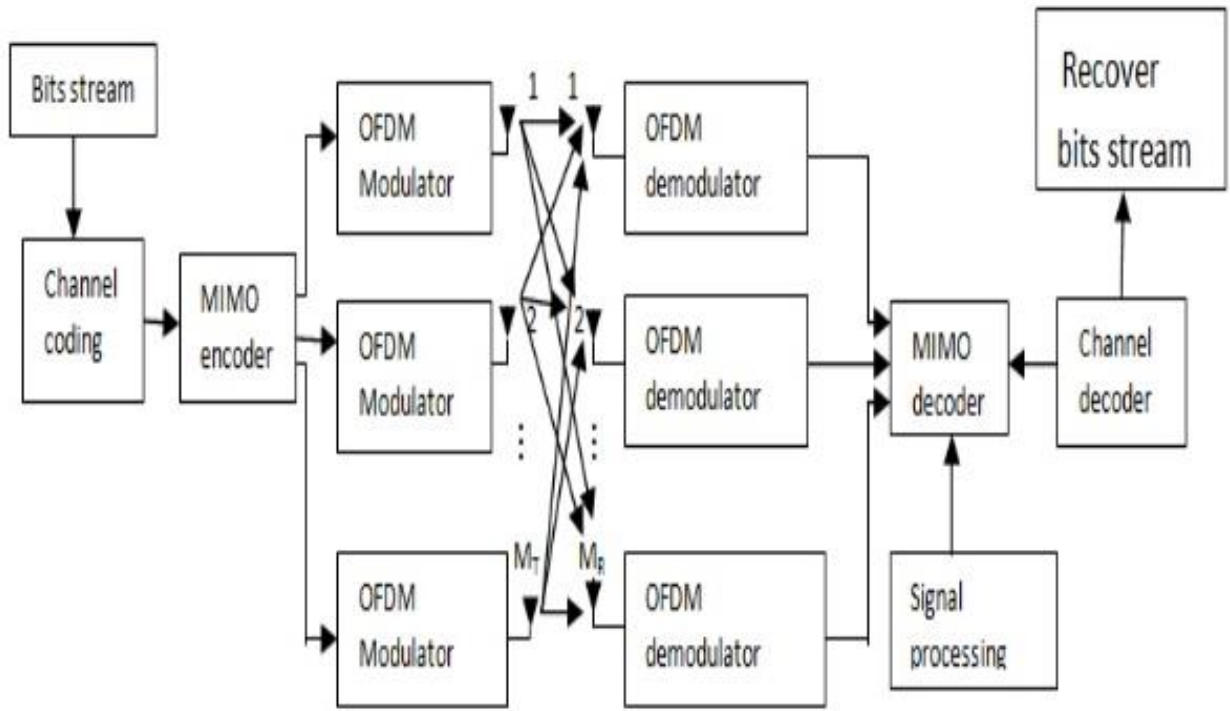

Figure3.1: OFDM System Model

## 3.2 Concept of FDM and OFDM

As shown in Figure 3.2(a), As in Frequency selective channels the coherence bandwidth is lesser than the data rate, so the Frequency division multiplexing (FDM) is being used. Frequency division multiplexing further divides the bandwidth of the channel into many subchannels, where each sub-channel has its own unique carrier frequency, and then multiplexing is being used to convert the stream of data into parallel streams of data having lower data rate. These parallel streams are then modulated on these unique carrier frequencies. To avoid the overlapping of these parallel signals at the demodulation, their carrier frequencies are placed at reasonably separate bands. One of the key features of FDM is that each subcarrier possesses a bandwidth less than the coherence bandwidth; thus only experiencing flat-fading and avoiding the frequency-selective fading. However, FDM requires generation of multiple carriers which demand high-precision local oscillators to be present both at the transmitting and receiving ends; making it a bit complex in terms of implementation. Furthermore, there is an inherent need for guard bands in the frequency spectrum between subcarriers to avoid the Adjacent-Channel Interference, which lowers the spectral efficiency of the FDM scheme.

In order to resolve the problems associated with FDM i.e. lower bandwidth efficiency and generation of multiple subcarriers, OFDM was proposed, which is based on modulating data symbols with orthogonal subcarriers, thus eliminating the need for guard bands. The subcarrier spacing is usually equal to symbol rate, which can be demodulated at the receiver using coherent techniques although the effective spectrum of each subcarrier overlaps with its neighbor. This setting of orthogonality with an optimal subcarrier spacing provides better spectral efficiency, much better than FDM. For example, if each symbol stream has a bandwidth of 2B, the FDM requires a bandwidth of 2L(B+GB), (GB represents guard band) while OFDM requires only total LB bandwidth, owing to the inherent orthogonality between the different symbols. This is shown in Figure 3.2 (b).

OFDM is widely used in various modern wireless communication systems such as the standard Wi-Fi Wireless LAN (802.11), WiMAX (802.16) and WiBro Wireless Broadband (802.20). Additionally, OFDM based multiple access technology is the standard multiple access scheme in LTE Broadband Mobile Technology.

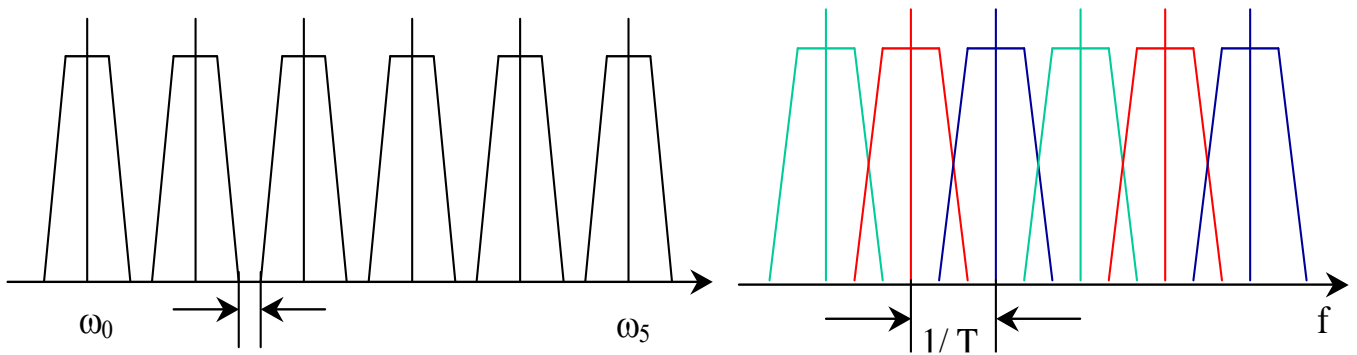

Figure: 3.2 FDM vs OFDM

OFDM assumes the data of kth symbol which is being transmitted can be expressed in complex as follows,

$$a(k) = z(k) + jy(k) \qquad (3.1)$$

At the transmitter output the signal can be expressed as follows:

$$\mathbf{x}(t) = \sum_{k=0}^{K-1} z(k)\cos(\omega_k t) + y(k)\sin(\omega_k t)$$

$$= \Re\left\{\sum_{k=0}^{K-1} (z_k + jy_k) e^{(j\omega_k t)}\right\} \tag{3.2}$$

$$= \Re\left\{\sum_{k=0}^{k-1} a_k e^{(jw_k t)}\right\}$$

Where $\omega_k = 2\pi f_k$ represents the sub-carrier frequency, $y(k)$ is the quadrature component and $z(k)$ is represented as in-phase component.

For calculating the real part we use the complex conjugate of equation 3.2:

$$\mathbf{x}(t) = \sum_{k=0}^{K-1} \frac{1}{2} \left\{ a_k e^{(j2\pi f_{ok} t)} + a_k^* e^{(-j2\pi f_o t)} \right\}$$

$$= \sum_{k=-(K-1)}^{K-1} \frac{1}{2} a_k e^{(j2\pi f_{ok} t)} \tag{3.3}$$

Where $k = 0 \ldots\ldots K-1$

$a_{-k} = a_k^*$, $d_0 = 0$, $f_{o(-k)} = -f_{ok}$, $f_{oo} = 0$,

$f_k = k\Delta f$, where $\Delta f$ is the carrier spacing

Equation 3.3 can be formalized as under by using Fourier coefficient Cn

$$C_n = \begin{cases} \frac{1}{2} a_n \text{ if } 1 \leq k \leq K-1 \\ \frac{1}{2} a_n^* \text{ if } -(k-1) \leq k \leq -1 \\ 0 \text{ if } k = 0 \end{cases} \tag{3.4}$$

Equation 3.3 can be expressed in a form close to DFT, as the Fourier coefficients of a real signal are complex conjugate symmetric,

$$\mathbf{x}(t) = \sum_{k=-(K-1)}^{K-1} C_k \cdot e^{j2\pi f_{ok} t} \tag{3.5}$$

Assuming the values of frequency of sub_channel carriers as $f_{ok} = kf_o$, where $k = 0 \ldots\ldots K-1$ and $f_o = 1/T$ represents the spacing between the sub_carriers which are reciprocal to the intervals of sub_channel signals, so the total bandwidth of one sided signal becomes equal to $B=(K-1)f_o$.

Now using the discretized time $t = i\Delta t$, where $\Delta t = 1/f_s$ is the reciprocal of the sampling frequency fs, (5) is can be written as:

$$\mathbf{x}(i\Delta t) = \sum_{k=-(K-1)}^{K-1} C_k \cdot e^{j2\pi f_{ok} i\Delta t} \tag{3.6}$$

The Nyquist criterion is satisfied when $f_s = 2(K-1)f_o$, where the sampling frequency $f_s$ is an integer multiple of sub-carrier spacing $f_o$ represented as $f_s = Mf_o$. Hence the spectrum of sampled signal repeats itself at integral multiples of the sampling frequency $f_s$ with a periodicity of $M = f_s/f_o$ samples. By exploiting the conjugate complex symmetry of the the Fourier coefficients $F_n$, we get:

$$C_n = \begin{cases} C_{k-M} = C^*_{M-k} & if \left(\frac{M}{2}+1\right) \leq k \leq M-1 \\ 0 & if K-1 \leq k \leq \frac{M}{2} \end{cases} \tag{3.7}$$

The idle band of the communication channel represented here as the frequency region $(K-1)f_o \leq f_k \leq \frac{M}{2}f_o$ in which delay distortion and amplitude is important. The real modulated signal after applying complex conjugate symmetry is as follows,

$$\mathbf{x}(i\Delta t) = \sum_{k=0}^{M-1} C_k \cdot e^{j\frac{2\pi}{M}ki}, i = 0 \cdots M-1 \tag{3.8}$$

The form is calculated by IFFT if the length of transformation M is the power of 2, this is known as standard IDFT.

## 3.3 OFDM-MIMO Implementation

Figure 3.3 shows the block diagram of an OFDM system. By looking at the figure we understand that OFDM basically consists of multiplexing and modulation. Splitting the single data stream into low-rate parallel data streams is done by using multiplexing. While for each sub-cariers, the mapping of data symbols as complexed value signal samples done by modulation. The key lead of an OFDM is that it removes the problem of ISI as well as it mitigates the frequency selective fading effect, due to the time delay spread of the channel impulse response.

In most of the systems, delay equalizer is being used to mitigate the effect of ISI due to delay spead of the channel. When delaying with a moving handset, as the conditions changes rapidly a major problem rises which is that how to use an adaptive delay equalizer. This imposes an important impact on the quality of the received data because adaptive equalizer will take much longer time to converge. A different approach to remove ISI in case of an OFDM is used which has an impotant advantage in implementing a communication scheme of high rate.

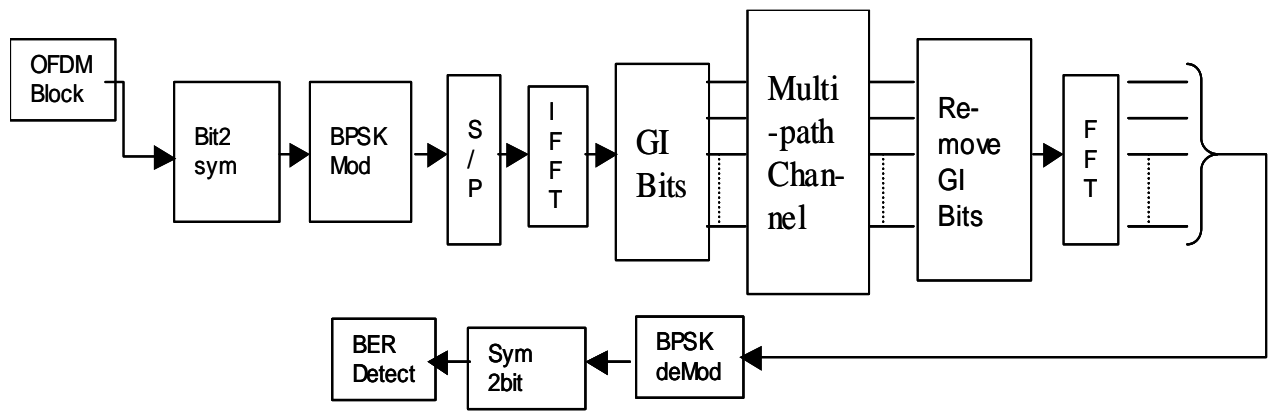

Figure3.3: Block Diagram of an OFDM System

For removing ISI in an OFDM system we introduces a time guard-band which should be longer than the delay spread of the channel (Tds) in between each transmitted OFDM signal. By using this, we ensure the stoppage of error travel from one symbol to the next, so the interference which corrupts one symbol is limited to that symbol only. But this kind of approach in signal carrier system is not implementable because it limits the data rate to 1/(2Tds).

Limiting the period of data bit will affect the BER and reduces the energy of every bit, because of guard interval in between the bits it will not be effective any more.

Because of multiplexing in an OFDM system the data rate has been divided into M no. of parallel low rate streams so, the minor duration of each symbol is extended to MTs because of the reduction factor M of data rate on each stream. For the choosen modulation scheme the symbol duration of data stream is Ts. The efficiency loss by using a guard interval will be much smaller than signal carrier case only if M is large enough such that MTs>>Tds. The parallel M no. of OFDM symbols are transmitted to achieve the data rate of the OFDM system. In practice, because of intolerable inter carrier interference the null signals are not used to represent guard band. To ensure the orthognonality between the sub-carriers the guard interval is a cyclic extension of the OFDM frame.

As compared to signal carrier system, the OFDM does not any degradation in performance as long as no guard interval is used. While on the other hand the BER performance will be reduced because energy has to waste during sending the cyclic performance when a guard interval is used. In Rayleigh fading channel it will suffer from ISI when used with Doppler. To solve this problem in this fading channel the delay spread should be smaller or equal to the length of an OFDM block as expressed below,

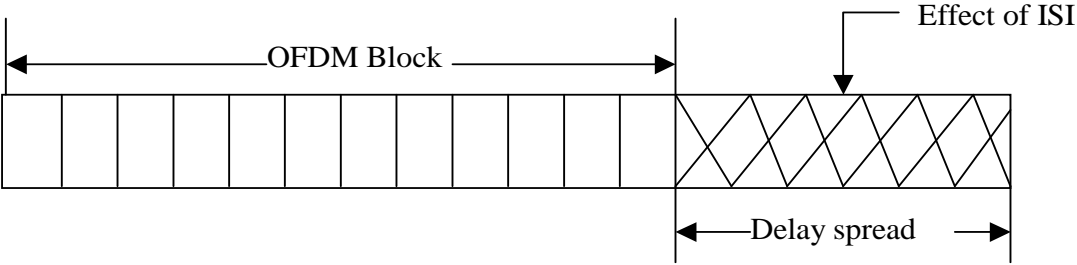

The maximum time delay that occurs is called delay spread of the signal in that particular environment. The delay spread can be large or shorter than the symbol time and in both cases different type of degradation occurs to the signal, as a result the delay spread changes with the change of an environment. It is not wrong to say that when the delay spread is less than one symbol period is considered to be a case of flat fading whereas if the delay spread is larger than one symbol period, the phenomenon is called frequency selective fading. Usually an OFDM system created works as is shown below in Figure.2.

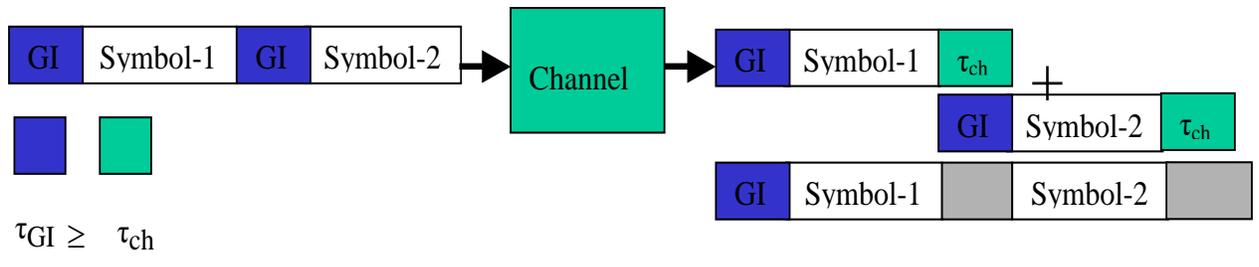

$\tau_{GI} \geq \tau_{ch}$

# CHAPTER 4

# STBC MIMO-OFDM system

In a mobile channel as we see some signal sub-carriers of the OFDM are more attenuated comparatively to other sub-carriers and data which travels on such sub-carriers could be lost if there exists a frequency selective fading. To overcome this problem the in-built diversity of STBC MIMO-OFDM system is totally exploited by introducing coding and inter-leaving through the sub-carriers. To reproduce the information effected by nulls of the frequency response of the channel, same connection can be used at receiver. This shows a deep connection among information which is being carried on various sub-carriers.

Schematic diagram of a codded OFDM-MIMO system can be seen in Figure4.2. In an OFDM system, basic principle is to introduce guard interval recognized as cyclic prefix (CP) which should be adequate to lodge delay spread of the channel. To turn the channel's action from linear to cyclic convolution on the transmitted signal, the CP is used. The ultimate purpose is to diagonalize the overall transfer function by using IFFT and FFT at transmitter and receiver respectively.

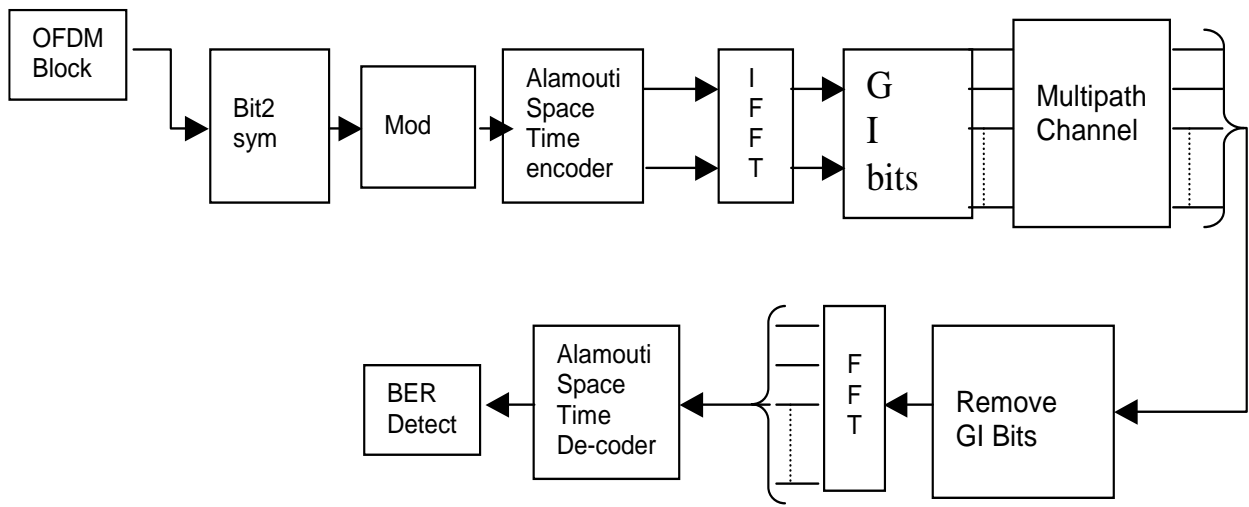

Figure 4.1: Detailed Block diagram of STBC MIMO-OFDM system

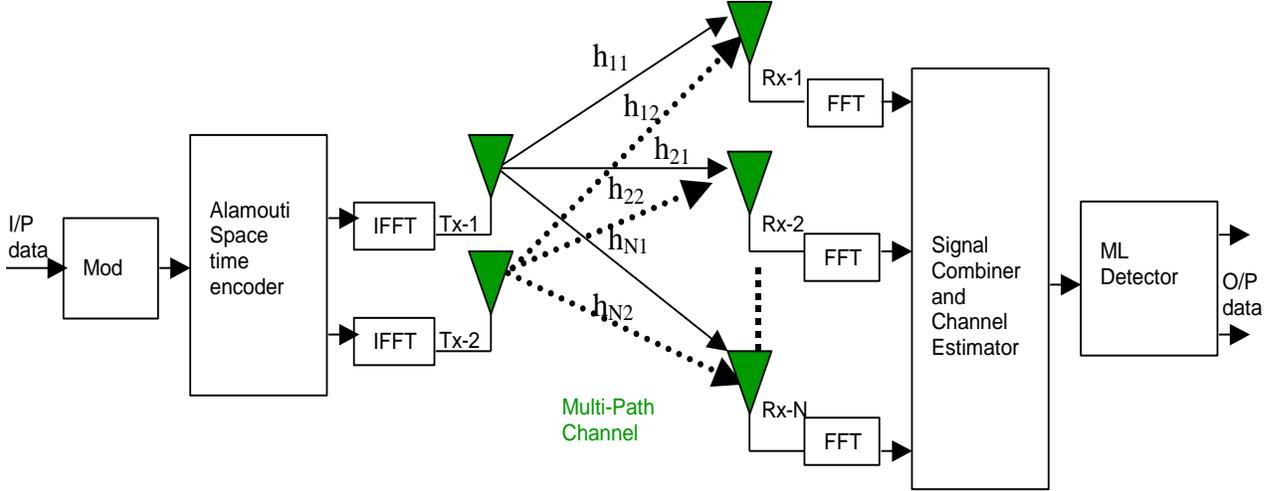

Figure 4.2: Schematic Diagram of STBC MIMO-OFDM System

Following equations express the OFDM transmitted signal for a OFDM-MIMO system with N no. of receive antenna and M no. of transmit antennas.

$$y(t) = \sum_{m=0}^{m-1} c(m)\cos(\omega_m t) - d(m)\sin(\omega_m t)$$

$$= \text{Re}\left(\sum_{m=0}^{m-1} b(m)\exp(j2\pi k \Delta f_m t).\exp(j2\pi f_o t)\right) \quad (4.1)$$

$$= \text{Re}\left(\tilde{B}(t)\exp(j2\pi f_o t)\right)$$

$d(m)$ and $c(m)$ are the imaginary and real parts of $M^{th}$ symbol $b(m)$ respectively, $\omega_m = 2\pi f_m$, $f_m = f_o + m\Delta f$, and $\tilde{B}(t)$ express the composite cover of the signal which is transmitted. At receiver the signal which is received by multiple antennas are the sum of the transmitted signal and ultimately the transmitted information is recovered by jointly process. At receiving antenna n the received signal, with the effect of noise $\mathbf{N}_{t,k}^{j}$, is shown mathematically as,

$$x_{t,k}^{j} = \sum_{i=1}^{M} \mathbf{H}_{j,i}^{t,k} \mathbf{x}_{t,k}^{i} + \mathbf{N}_{t,k}^{j} \quad (4.2)$$

Where $\mathbf{H}_{j,i}^{t,k}$, defines the coefficient of multipath quasi-static channel between M no. of transmitter and N no. of receiver and $\mathbf{N}_{t,k}^{j}$ express the effect of noise.

There exist various analogue damages like phase noise and I/Q disparity present in MIMO-OFDM system which limit and considerably degrade performance of the communication systems. The above discussed damages addressed properly in case where the performance of a MIMO-OFDM system is corrupted, when the sub-carriers lose their orthogonal property because of these impairments but at the same time by cancelling these impairments, the performance of the system is improved.

## 4.1 Receiver & Channel Parts:

Let's say the signal $\tilde{\mathbf{x}}_n[k]$ which is being transmitted through the linear and time variant wireless channel $h(t,\tau)$ which has no noise factor. Where $h[k]$ is the time sampled version of the channel $h(t,\tau)$, then the output would be as described below mathematically:

$$\tilde{y}_n[k] = \sum_{l=0}^{N+L-1} \tilde{\mathbf{x}}_n[l] h[k-l] \qquad (4.3)$$

$$k = 0, 1, 2, 3, \ldots, N+L-1$$

At the receiving end the reverse process is being done. $\tilde{y}_n[k]$ is the signal which is being received and it has $N+L-1$ number of samples, In order to use the properties of Fourier transform efficiently, it is required to drop last L number of samples from the received signal before demodulation process and then removal of guard interval bits is done. IFFT was used for the modulation purpose so it is simple to demodulate the received signal by using FFT. At the input of modulator the N number of samples are found out, so demodulating the received signal $\tilde{y}_n[m]$ by using FFT, we have:

$$\tilde{Y}_n[k] = \sum_{l=0}^{N-1} \tilde{y}_n[l] \exp\left(-2\pi j k \frac{l}{N}\right) \qquad (4.4)$$

$$\tilde{Y}_n[k] = \sum_{l=0}^{N-1} \left( \sum_{m=0}^{N+L-1} \tilde{x}_n[k] h[l-k] \right) \exp\left(-2\pi j k \frac{l}{N}\right) \qquad (4.5)$$

$$\tilde{Y}_n[k] = \left( \sum_{m=0}^{N+L-1} \tilde{x}_n[k] \right) \sum_{l=0}^{N-1} h[l-k] \exp\left(-2\pi j k \frac{l}{N}\right) \qquad (4.6)$$

$$\tilde{Y}_n[k] = \left( \sum_{m=0}^{N+L-1} \tilde{x}_n[k] \exp\left(-2\pi j k \frac{l}{N}\right) \right) \Im(h[k]) \qquad (4.7)$$

If we limit the index of summation from m = 0 to k = N-1 it is same as to drop last L number of samples from the received sampled signal $\tilde{y}_n[k]$. So the signal after applying FFT is given as under:

$$\tilde{Y}_n[k] = \left(\sum_{m=0}^{N-1} \tilde{x}_n[k] \exp\left(-2\pi j k \frac{l}{N}\right)\right) \Im(h[k]) \tag{4.8}$$

$$\tilde{Y}_n[k] = \Im(x_n[k]) \Im(h[k]) \tag{4.9}$$

$$\tilde{Y}_n[k] = X_n[k] H[k] \tag{4.10}$$

For getting back the original transmitted signal we divide the output signal with the frequency response of the channel. There is no requirement of equalization in this modulation scheme so the data samples are then aligned as the original data's size.

## 4.2 Channel estimation for the STBC OFDM-MIMO Systems

In a coherent communication model the estimation of a channel plays a vital role. For MIMO system this is especially true because in this system multiple inputs and multiple outputs can be achieved by using multiple antennas. As in MIMO system multipath channel provides the space diversity so, it is required to have knowledge of the characteristics of all channel for each multipath, so that by using interference cancellation techniques the interference introduced by co-channel link might be removed successfully. In this paper a realistic approach to estimation of channel has been taken in which we are using pilot symbols. In this approach, receiver knows all the data symbols of an OFDM frame which were sent. During the transmission channel corrupts the transmitted pilot symbols, but as the receiver has knowledge of the sequence of symbols, it is possible to re-construct the channel state information (CSI). By using the above said approach phase and amplitude problems can be easily resolved for each sub-carrier in the frequency domain. In this method, pilot symbols are the raw symbols which are not the decoded bits but have the knowledge of channel state information (CSI) so, these symbols should not be inter-leaved. After the estimation of channel by frame of channel bits, the left over frames contains the symbols of payload only. Channel tends to change after estimation of the channel is done, because channel shows time varying characteristics. This time varying characteristics can be set by using channel coherence time which is also linked with the mobile user's velocity. Therefore whenever the time period is smaller than the channel coherence time, a channel re-estimate is done periodically by re-transmitting the pilot symbols. The same technique is being used to generate the results of figure 4.4 to figure 4.9.

As in coded MIMO-OFDM system, the problem arises when a signal is transmitted simultaneously on various channels. It is important that estimation of the characteristics of every single path is made simultaneously. To achieve this a new method is derived to estimate the channel in which a unique code of pilot symbols is associated with each transmitter which is orthogonal to all other codes associated with other transmitter elements forming a MIMO system. These unique codes are drawn from Hadamard orthogonal sequences which are the rows of the Hadamard matrix. Frame length of OFDM should be same as the Hadamard sequence length. BPSK modulation is used for pilot symbols to reconstruct the CSI from the Hadamard sequence. We implemented this approach in MATLAB and found it extremely well for the coded OFDM-MIMO systems.

Recursive Least Squares (RLS) algorithm for channel estimation can be used for channel estimation of STBC OFDM-MIMO systems. In which a transfer function of the channel is made for frequency domain by using the relationship of inputs and outputs then inverse fast Fourier transform (IFFT) is applied on transfer function to get the matrix of channel parameter in time domain. Later in order to estimate the channel, RLS/LMS adaptive filter is used, finally to get the matrix in frequency domain of the channel transfer function fast Fourier Transform (FFT) is used.

## 4.3 Communication System Selection Criterion

In wideband communication system, the radio channel is time varying and frequency selective, therefore OFDM signal demodulation is done after the estimation of the channel. It is important to define channel matrix before using communication system in a specific outdoor environment, because to decide which technique of communication is used, either MIMO, Beam forming, Opportunistic Communication, coded OFDM or OFDM the information of channel matrix "H" is required.

### 4.3.1 Method of Channel Matrix ($\mathbf{H}$) Selection
- o If any of the rows contains not the same power than row with the lower power is deleted but the row with the higher power is selected.

- In case rows of channel matrix (H) are not alike, then with the highest correlation two rows are chosen. One with the highest correlation is selected while the second one is deleted.
- If any of the two rows of channel matrix (H) contains the same information then one of them is selected and the other one is deleted.

Therefore the channel matrix $\tilde{\mathbf{H}}$ which is being estimated is selected with maximum uncorrelated rows and rows with the maximum power.

$\mathbf{H} * \mathbf{H}^H = 1$ Defines the fully correlation of $\mathbf{H}$, whereas $\mathbf{H} * \mathbf{H}^H = 0$ defines the un-correlation of $\mathbf{H}$.

To indicate the correlation matrix $\mathbf{H}$, the channel coefficients are being generated. Line of sight (LOS) environment dominates in case of flat fading and it happens when estimated channel coefficients are being correlated so, opportunistic communication system, MIMO or Beam forming techniques could be used.

Opportunistic communication technique can applied in this case because fair sharing of resources will be required in case a null is formed, and it can be achieved by using a switch. MIMO technique in which antenna elements are adequately spaced a part be entirely influenced by spatial correlation of antenna elements, stations with narrow space properties. It can give us an adequate performance only when that station is located at where de-correlation is very little.

While on the other hand SDMA technique is used in Beam forming which relies on estimation of angle of arrival (AoA) and it is most effective in case of flat fading.

A frequency selective channel is occurred when estimated channel coefficients are un-correlated, then it shows the frequency selectivity property which is also known as delay spread. Inter symbol interference (ISI) occurs due to the result of delay-spread and it causes a serious degradation in performance. In this case to remove the effect of ISI, OFDM or COFDM systems can be used as these techniques offer greater spectral efficiency and remove the need for equalization. When delay spread is lengthier than cyclic prefix, ICI is also created. Frequency offset between transmitter and receiver defines the time variation of channel only when channel diverges between symbol periods of OFDM. When fast fading occurs but BER does not increase with change in SNR, the reason is that tracking of channel estimate is poor. Increase in Doppler spread causes an increase in BER because of severe ICI.

**4.3.2 Diversity**

In a fading environment to get multipath diversity, multiple antennas are to be used at transmitter and receiver. Multiple spatial channels are formed by introducing multiple antennas, and it is improbable that all the channels will fade concurrently.

Transmitting the signal through more than one antennas where space, frequency or time dimensions are independently faded and by appropriate joining of the signal at the receiver can give us a diversity gain. Spatial multiplexing can give us far better results as compared to frequency or time diversity, as it does no suffer expenses in bandwidth or time of the transmission. Even if channel behavior is unknown at the transmitting end, still space time codes with the help of multiple antennas, comprehends spatial diversity gain into the system. Two sources of diversity such as frequency and spatial diversity are available in MIMO channels with frequency selective environment to be explored using STBC codes.

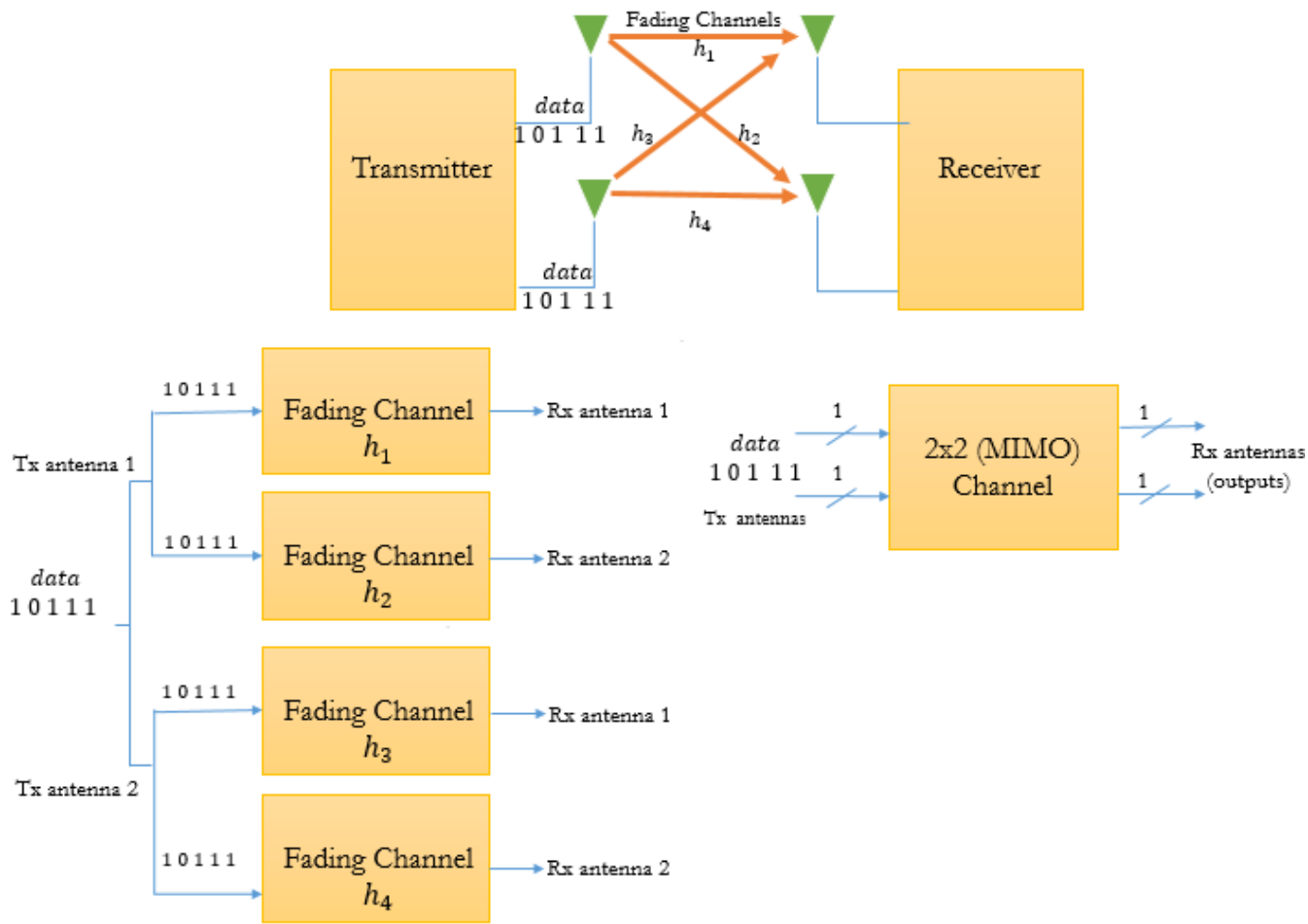

Figure 4.3: Multi Input Multi Output (MIMO) with Diversity

## 4.4 Implementation of STBC MIMO-OFDM Systems

We have implemented two coded OFDM-MIMO systems in MATLAB using Spatial Channel Model & Rayleigh Fading Channel in frequency selective faded environment. Table 4.1 below contains the design parameters for the simulation of Spatial Channel Model and Rayleigh Fading Channel well-matched to specification requirements given by standard of IEEE 802.20.

| *Design Parameters* | *Values* | |
|---|---|---|
| Channels Used | Rayleigh Fading Channel | SCM |
| FFT bins | 64 | 64 |
| No. of blocks | 20,000 | 20,000 |
| No. of Symbols per block | 64 | 64 |
| Spacing between antennas | $\lambda/2$ | 0.5 |
| Environment used | frequency selective fading/Flat fading | frequency selective fading/Flat fading |
| Length for tracing back | 5* Constraint length (5*3) | 5* Constraint length (5*3) |
| Carrier frequency (Hz) | 2.4 GHz | 2.4 GHz |
| Signal to Noise Ratio (SNR) | 1: 35 dB | 1: 35 dB |
| Doppler frequency (Hz) | 100/141.6 Hz (63mph) | 10/50/100/141.6 Hz (63mph) |
| Coding rate | ½ of Convolutional Rate | ½ of Convolutional Rate |
| Bandwidth (Hz) | 10 MHz | 10 MHz |
| Sampling time/symbol (sec) | 0.2 μsec | 0.2 μsec |
| No. of Paths/Link | 2/3/4/5 | 2 |
| Modulation type | QAM/PSK | QAM/PSK |
| Maximum delay (sec) | 5μsec | 5μsec |
| No. of Sub-paths/path | - | 20 |
| Constraint length | 3 | 3 |
| Modulation order (M) | 2,4,16 | 2,4,16 |
| Guard Interval (GI) bits | 10 | 10 |

Table 4.1: Parameters for simulating STBC MIMO-OFDM Systems

## 4.5 Implementation of Rayleigh Fading Channel (STBC MIMO-OFDM system)

### 4.5.1 CASE-1

We have implemented a coded MIMO-OFDM system that considerably rises the bit rate while reducing inter-symbol and co-channel interference in time varying multipath MATLAB Rayleigh fading channel model. To explore the effect of diversity, BER vs SNR simulations has been implemented using Almouti space time coding scheme with various no. of antennas in MATLAB. The simulation results are shown in figure 4.4 to figure 4.7.

In case-1 a (2x2) coded MIMO-OFDM system is used for Rayleigh faded channel in frequency selective environment. Here in case of 2-paths the frequency inter-leaver is used to enhance the BER performance of the above said system. To see the influence of the multipath the same system is then instigated with 3 and 4 paths.

Design parameters for the simulation of this system are also available in Table4.1.

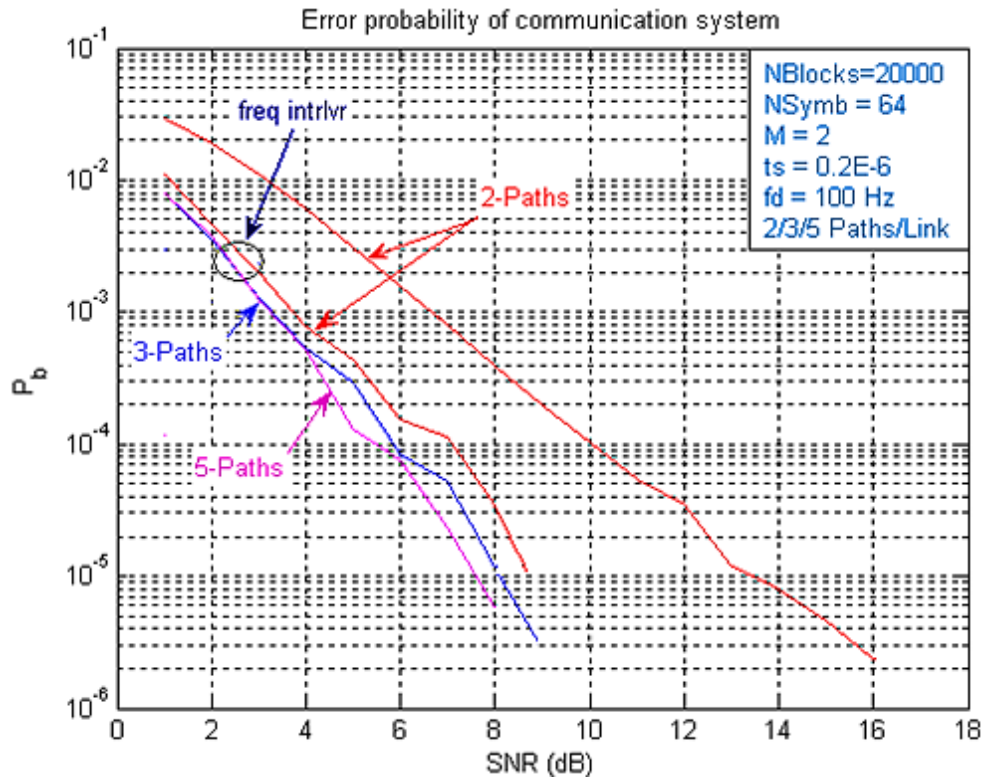

Figure 4.4: Effect of Multi-Paths on BER Performance

**4.5.1.1 Results Analysis**

BER performance of two paths Rayleigh Fading Channel can be seen from the simulation results of figure 4.4. The channel is encoded with guard interval bits and the environment is frequency selective. The curve specifies BER of $10^{-5}$ at SNR of 13dBs.

It is clear from the results that while using frequency inter-leaver along with Viterbi-decoder, it can give us the maximum benefits from frequency diversity. As it is witnessed that there is a substantial improvement in BER performance that the same BER of $10^{-5}$ can now be achieved at SNR of 9dBs as compared to 13dBs without using inter-leaver. So, it can improve the SNR by 4dBs at same BER rate.

While on the other hand Figure 4.4 also give us the comparison of the effect introduced by the multipath. By looking at the result it is clear that by using more no. of paths as shown in magneta and blue colors the BER performance is enhanced. This result also verifies the OFDM system's property that it diminishes the multipath effect. By using time and frequency inter-leavers the BER performance with 3 and 4 multi-paths can improve further.

**4.5.2 CASE-2**

In case-2 two (2x2) and (2x4) coded MIMO-OFDM systems is used for simulation of Rayleigh faded channel in frequency selective environment. In this case frequency inter-leaver along with Viterbi decoder is used to further enhance the BER performance of the system. At the end performance of both the systems (2x2) COFDM-MIMO and (2x4) COFDM-MIMO is compared. Design parameters for the simulation of this system are also available in Table 4.1.

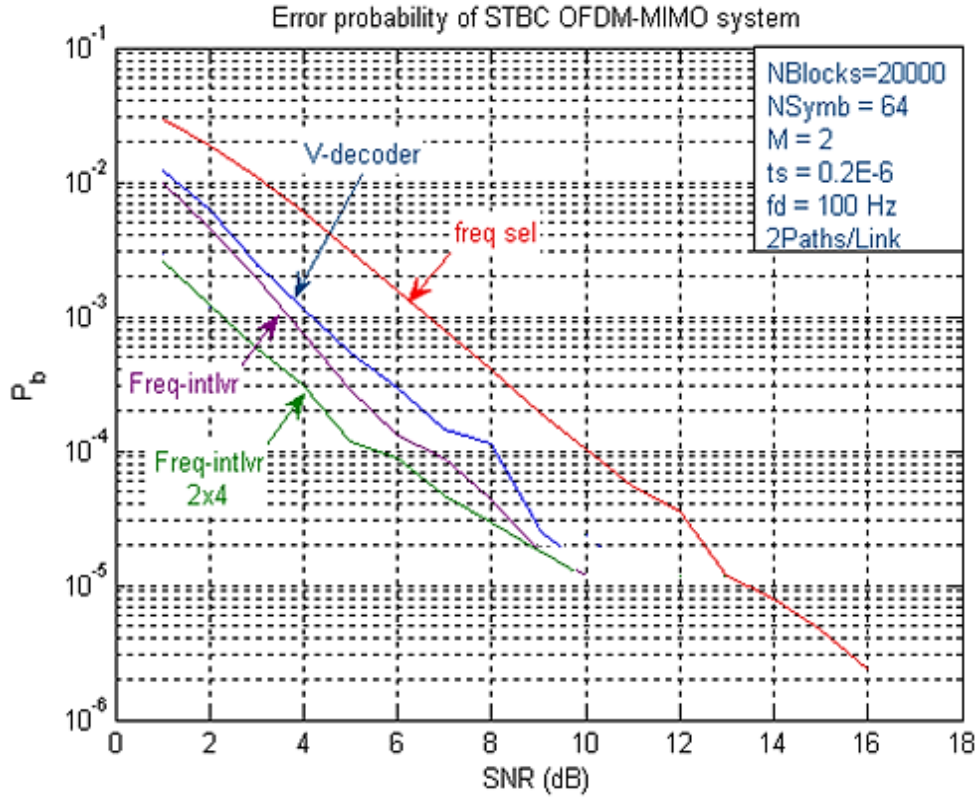

Figure 4.5: Diversity effect on BER Performance using frequency inter-leaver and Viterbi decoder

**4.5.2.1 Results Analysis**

Frequency inter-leaver is used alongside with Viterbi decoder to gain the maximum advantages from frequency diversity .Form figure 4.5 the following statements are being made,

  i)  In Rayleigh Fading Channel without using inter-leaver or Viterbi decoder at SNR of 13dBs the $10^{-5}$ BER is being achieved.
  ii) While using Viterbi decoder alone the same BER is achieved at SNR of 9dBs.

      iii)      By using frequency interleaving along with Viterbi decoder the same BER is now achieved at SNR of 9dBs, thus giving us an improvement in SNR by 3-4dBs.

While on the other hand by using 4-antennas at receiving end (2x4 system) with inter-leavers, further improvement in BER performance is achieved as compared to the system in which only two antennas were used at receiving end (2x2 system). In (2x4) COFDM MIMO, same BER is achieved at SNR of 8dBs thus giving an improvement in SNR of 1dB.

## 4.6 Implementation of Spatial Channel Model (STBC MIMO-OFDM system)

### 4.6.1 CASE-1

In this case we have implemented a dynamic and realistic spatial channel model using STBC MIMO-OFDM technology in frequency selective and flat fading environment. This system decreases the inter symbol and co-channel interference, while considerably increases the bit rate. To examine the effect of inter-leavers, modulation schemes and Doppler Effect, BER vs SNR results has been implemented by using Almouti space time coding scheme with various system parameters in MATLAB. These results are given in figure 4.6 to figure 4.8.

Design parameters for the simulation of this system are also available in Table 4.1.

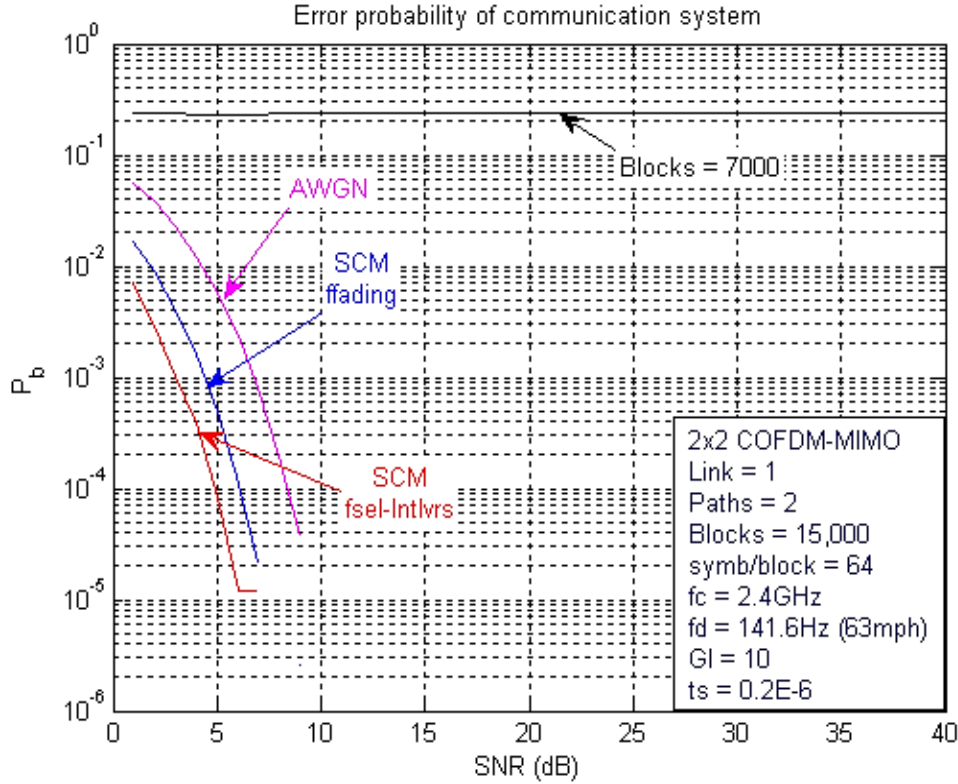

Figure 4.6: BER vs SNR for coded OFDM-MIMO SCM System

### 4.6.1.1 Results Analysis

From the results of figure 4.6, it is clearly seen that in SCM system model BER performance for frequency selective faded environment is much better as compared to flat fading environment. In case of 2 paths per link, flat fading environment can give BER of 10-4 at SNR of 6dBs but the frequency selective fading environment can give the same BER at SNR of 5dBs thus giving an improvement in SNR of 1dB. This improvement is due to applying the frequency/time inter-leavers.

By exploiting the effect of diversity and use of MIMO system these results are much better than the novel AWGN channel.

### 4.6.2 CASE-2

In this case we have implemented coded MIMO-OFDM system in frequency selective faded environment using realistic spatial channel model with various modulation techniques. To analyze the BER performance simulation results are generated by using QAM and PSK modulation techniques where M (modulation order) is taken as 2,4,16.

Design parameters for simulation of this system are also available in Table 4.1.

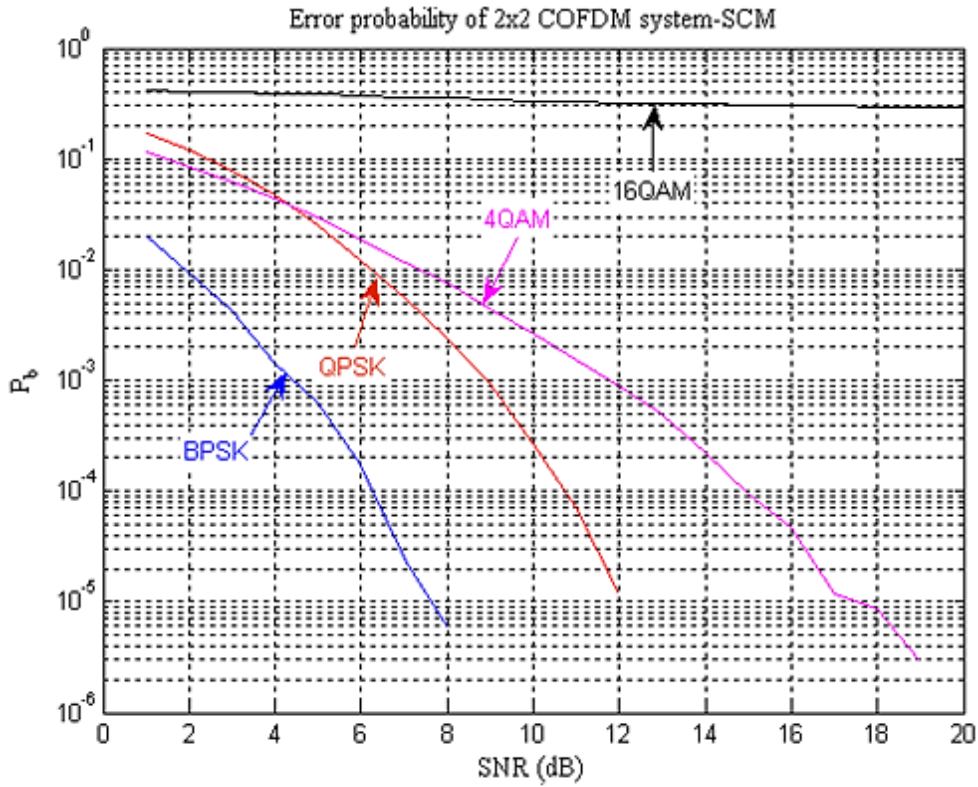

Figure 4.7: Comparison of different Modulation Schemes using SCM

### 4.6.2.1 Results Analysis

Figure 4.7 above shows the simulations for various modulation techniques. It can be seen from figure that BER performance of PSK modulation technique is enhanced than QAM modulation technique, because constellation points in case of QAM resides on square shape. Energy of a modulation technique can be calculated by the power of constellation points. In case of 4-QAM and QPSK modulation the power of constellation points is same while in 16-QAM it is different. As the constellation of 16-QAM modulation expands and points are spread on larger area, so those points which lie on the inner circle has less power as compared to the points on the outer circle of the constellation diagram. While on the other hand in 16-PSK all the constellation lies on the same circle therefore power of every point is same. In case of deep null in QAM modulation technique when likely points are nearer to origin (0,0) axis can't see the points of constellation on the wider circle therefore resulting in large error in data, this scenario is visible with a black flat line in figure-6 above.

In 4-QAM modulation scenario the BER performance is good when SNR is low but as the SNR increases BER performance is damaged because of poor channel estimation, sudden change in characteristics of channel due to Doppler effect and multipath effect.

By analyzing and comparing the BER performance of 4-QAM and QPSK, it is seen in 4-QAM BER of $10^{-5}$ is achieved at SNR of 17dBs while in QPSK the same BER of $10^{-5}$ is now achieved at SNR of 12dBs, therefore giving an improvement of 5dBs. In case of BPSK modulation technique the same BER is achieved at SNR of 7-8dBs.

### 4.6.3 CASE-3

In this case again we have implemented coded MIMO-OFDM system in frequency selective faded environment using realistic spatial channel model with various modulation techniques. The simulations are performed to check the Doppler Effect using QAM and PSK modulation techniques.

Design parameters for simulation of this system are also available in Table 4.1.

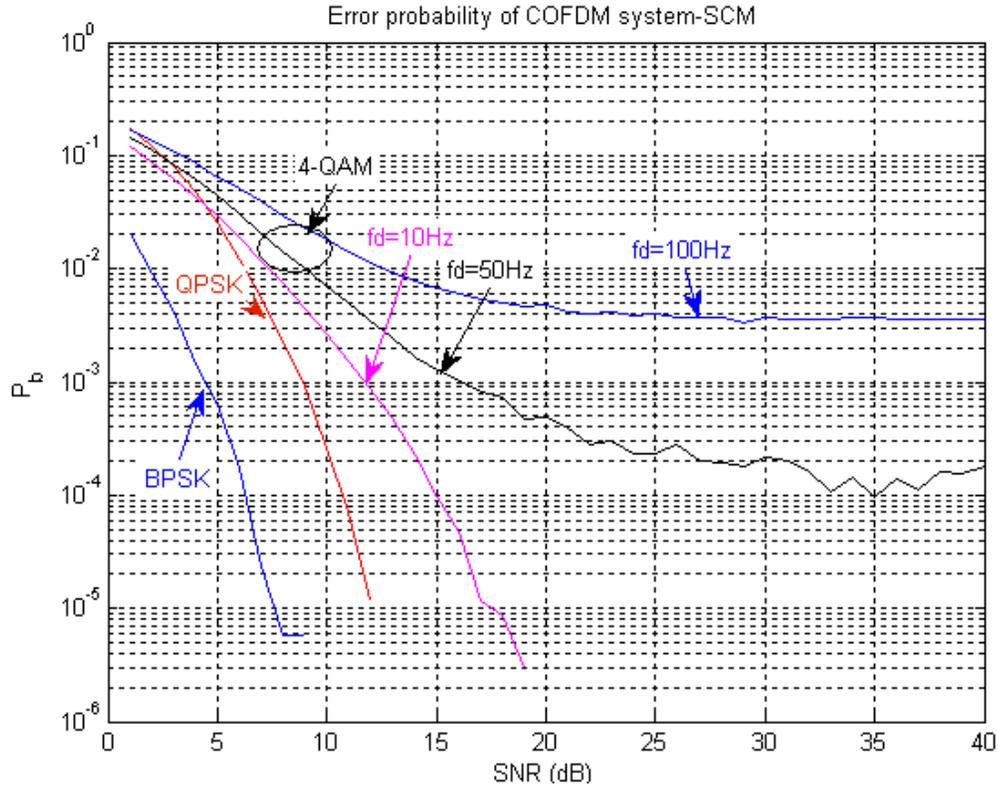

Figure 4.8: Doppler Effect on SCM using MIMO-OFDM

**4.6.3.1 Results Analysis**

In this case the simulations results re being generated to see the Doppler Effect on BER performance using various modulation techniques. QPSK shows better results as compared to QAM when Hz of Doppler is considered because of distribution of power of constellation points as discussed in previous case.

In QAM modulation technique a there is a degradation in BER performance as the Doppler because of fast change in characteristics of the channel.

BER performance is only compatible to Doppler of Hz at low SNR and then it is faded. Because of multipath environment and sudden change in channel characteristics, there is a irreducible BER floor is accomplished at high SNR.

# CHAPTER 5

# CONCLUSION AND FUTURE WORK

## 5.1 CONCLUSION

In this document, first a simulation environment for outdoor channel model for MU-MIMO was created in MATLAB. Three different scenarios were considered: urban, micro, and urban macro. Simulation results were obtained for these scenarios for input link, antenna, and SCM parameters. Simulation results demonstrate that user parameters – AoA, user direction and distance between user and AP - in a MU-MIMO system in an outdoor environment that may fall in any of above scenarios can be accurately extracted using the proposed adaptive algorithm.

Secondly We have implemented COFDM-MIMO system on two different channels (Rayleigh Fading Channel Model & Spatial Channel Model), as we analyse various cases in which Almouti Space time coding, Viterbi Decoding and Frequency Inter-leavers were used to increase the BER performance, we have come to a conclusion that SCM can be more efficient in the sense of BER vs SNR as compared to Rayleigh Fading Channel Model. In SCM COFDM-MIMO system at SNR of 6dB BER of 10-5 is achieved in case of 2 paths per link while on the other hand in case of multipath Rayleigh Faded Channel Model at SNR of 9 dB BER of 10-5 was achieved for (2x2) system.

Another effect which can be seen from the cases discussed above as we introduce the effect of diversity by increasing the order of diversity at receiving end we can achieve enhanced BER vs SNR. It is shown in figure 4, for (2x2) system at SNR of 9dB BER of 10-5 is achieved while in (2x4) system in which 4 antennas are used at receiving end at SNR of 8dB BER of 10-5 was achieved.

## 5.2 FUTURE WORK

- Implementation of Massive MIMO (Multiple Input Multiple Output) technology on Spatial Channel Model (SCM).
- Implementation of proposed MU-MIMO technology on 5$^{th}$ Generation Mobile Network.

- Enhancement of BER vs SNR using several other Space Time Coding techniques.
- Implementation of advanced Spatial Multiplexing techniques to increase data rate.

# Bibliography

[1] Paleerat Wongchampa, Nakhon Ratchasima, Monthippa Uthansakul. "Data rate and throughput enhancement base on IEEE802.11n standard employing multiple antenna elements". Electrical Engineering/Electronics, Computer, Telecommunications and Information Technology (ECTI-CON),11th International Conference (2014) : 1-4

[2] George Jongren, Bo Goransson, Lei Wan. "Mode Switching Between SU-MIMO and MU-MIMO" US Patent no: US 20100322330 A1, Dec 23, 2010

[3] Liu, Lingjia, et al. "Downlink mimo in lte-advanced: SU-MIMO vs. MU-MIMO." Communications Magazine, IEEE 50.2 (2012): 140-147.

[4] Syed Waqas Haider Shah, Shahzad Amin Sheikh, Khalid Iqbal, "An adaptive algorithm for mu-mimo using spatial channel model". International Journal of Engineering. (volume:10, issue:1) (2016), 1-13

[5] Yoshihide Nomura, Kazuo Mori, Hideo Kobayashi. "Efficient Frame Aggregation with Frame Size Adaptation for Next Generation MU-MIMO WLANs" Next Generation Mobile Applications, Services and Technologies, 2015 9th International Conference by IEEE.(2015). 288 – 293

[6] Christoph Studer, Erik G. Larsson "PAR-Aware Large-Scale Multi-User MIMO-OFDM Downlink". IEEE Journal on Selected Areas in Communications (Volume:31 , Issue: 2 ) (2013), 300-313

[7] Duplicy, Jonathan, et al. "Mu-mimo in lte systems." EURASIP Journal on Wireless Communications and Networking 2011.1 (2011): 496763.

[8] Ghassan Dahman, Jose Flordelis, Fredrik Tufvesson." Experimental evaluation of the effect of BS antenna inter element spacing on MU-MIMO separation". Communications (ICC), 2015 IEEE International Conference. (2015) 1685 – 1690.

[9] Erik Larsson, Ove Edfors, Fredrik Tufvesson, T. Marzetta. "Massive MIMO for next generation wireless systems" IEEE Communications Magazine (Volume:52 , Issue: 2 ) (2014): 186 – 195

[10] Bölcskei, Helmut, David Gesbert, and Arogyaswami J. Paulraj. "On the capacity of OFDM based spatial multiplexing systems." Communications, IEEE Transactions on 50.2 (2002): 225-234.

[11] Li, Qinghua, et al. "MIMO techniques in WiMAX and LTE: a feature overview." Communications Magazine, IEEE 48.5 (2010): 86-92.

[12] Sicong Liu, Fang Yang, Jian Song, Jianqi Li. "An optimized time-frequency interleaving scheme for OFDM-based power line communication systems". 2014 IEEE International Conference on Communications (ICC) Sydney (10-14 June 2014), 4137-4142

[13] Zafar Iqbal, Saeid Nooshabadi. "Effects of channel coding and interleaving in MIMO-OFDM systems". 2011 IEEE 54th International Midwest Symposium on Circuits and Systems (MWSCAS) Seoul, (7-10 Aug 2011), 1-4

[14] Andrews, Jeffrey G., Wan Choi, and Robert W. Heath. "Overcoming interference in spatial multiplexing MIMO cellular networks." Wireless Communications, IEEE 14.6 (2007): 95-104.

[15] Gan, Ying Hung, Cong Ling, and Wai Ho Mow. "Complex lattice reduction algorithm for low complexity full-diversity MIMO detection." Signal Processing, IEEE Transactions on 57.7 (2009): 2701-2710.

[16] Ramprashad, Sean A., Giuseppe Caire, and Haralabos C. Papadopoulos. "Cellular and network MIMO architectures: MU-MIMO spectral efficiency and costs of channel state information." Signals, Systems and Computers, 2009 Conference Record of the Forty-Third Asilomar Conference on. IEEE, 2009.

# Completion Certificate

"It is certified that the contents of thesis document titled "*A Technique for Multi-User MIMO using Spatial Channel Model for out-door environments*" submitted by Mr. Syed Waqas Haider Shah Registration No. NUST201464406MCEME35014F have been found satisfactory for the requirement of degree".

Thesis Advisor: ______________________

(Dr. Shahzad Amin Sheikh)